\documentclass[twocolumn]{aastex631}

\usepackage{amsmath} 
\usepackage[T1]{fontenc} 

\begin{document}

\title{exoALMA. XVI. Predicting Signatures of Large-scale Turbulence in Protoplanetary Disks}


\correspondingauthor{Marcelo Barraza-Alfaro}
\email{mbarraza@mit.edu}

\author[0000-0001-6378-7873]{Marcelo Barraza-Alfaro}
\affiliation{Department of Earth, Atmospheric, and Planetary Sciences, Massachusetts Institute of Technology, Cambridge, MA 02139, USA}

\author[0000-0002-9298-3029]{Mario Flock}
\affiliation{Max-Planck Institute for Astronomy (MPIA), Königstuhl 17, 69117 Heidelberg, Germany}

\author[0009-0001-2471-0336]{William B\'ethune}
\affiliation{DPHY, ONERA, Université Paris-Saclay, 91120, Palaiseau, France}

\author[0000-0003-1534-5186]{Richard Teague}
\affiliation{Department of Earth, Atmospheric, and Planetary Sciences, Massachusetts Institute of Technology, Cambridge, MA 02139, USA}

\author[0000-0001-7258-770X]{Jaehan Bae}
\affiliation{Department of Astronomy, University of Florida, Gainesville, FL 32611, USA}

\author[0000-0002-7695-7605]{Myriam Benisty}
\affiliation{Universit\'{e} C\^{o}te d'Azur, Observatoire de la C\^{o}te d'Azur, CNRS, Laboratoire Lagrange, France}
\affiliation{Max-Planck Institute for Astronomy (MPIA), Königstuhl 17, 69117 Heidelberg, Germany}

\author[0000-0002-2700-9676]{Gianni Cataldi} 
\affiliation{National Astronomical Observatory of Japan, 2-21-1 Osawa, Mitaka, Tokyo 181-8588, Japan}

\author[0000-0003-2045-2154]{Pietro Curone} 
\affiliation{Dipartimento di Fisica, Universit\`a degli Studi di Milano, Via Celoria 16, 20133 Milano, Italy}
\affiliation{Departamento de Astronom\'ia, Universidad de Chile, Camino El Observatorio 1515, Las Condes, Santiago, Chile}

\author[0000-0002-1483-8811]{Ian Czekala}
\affiliation{School of Physics \& Astronomy, University of St. Andrews, North Haugh, St. Andrews KY16 9SS, UK}

\author[0000-0003-4689-2684]{Stefano Facchini}
\affiliation{Dipartimento di Fisica, Universit\`a degli Studi di Milano, Via Celoria 16, 20133 Milano, Italy}

\author[0000-0003-4679-4072]{Daniele Fasano} 
\affiliation{Universit\'{e} C\^{o}te d'Azur, Observatoire de la C\^{o}te d'Azur, CNRS, Laboratoire Lagrange, France}

\author[0000-0003-1117-9213]{Misato Fukagawa} 
\affiliation{National Astronomical Observatory of Japan, 2-21-1 Osawa, Mitaka, Tokyo 181-8588, Japan}

\author[0000-0002-5503-5476]{Maria Galloway-Sprietsma}
\affiliation{Department of Astronomy, University of Florida, Gainesville, FL 32611, USA}

\author[0000-0002-5910-4598]{Himanshi Garg}
\affiliation{School of Physics and Astronomy, Monash University, Clayton VIC 3800, Australia}

\author[0000-0002-8138-0425]{Cassandra Hall} 
\affiliation{Department of Physics and Astronomy, The University of Georgia, Athens, GA 30602, USA}
\affiliation{Center for Simulational Physics, The University of Georgia, Athens, GA 30602, USA}
\affiliation{Institute for Artificial Intelligence, The University of Georgia, Athens, GA, 30602, USA}

\author[0000-0001-6947-6072]{Jane Huang} 
\affiliation{Department of Astronomy, Columbia University, 538 W. 120th Street, Pupin Hall, New York, NY, USA}

\author[0000-0003-1008-1142]{John~D.~Ilee} 
\affiliation{School of Physics and Astronomy, University of Leeds, Leeds, UK, LS2 9JT}

\author[0000-0001-8446-3026]{Andr\'es F. Izquierdo} 
\affiliation{Department of Astronomy, University of Florida, Gainesville, FL 32611, USA}
\affiliation{Leiden Observatory, Leiden University, P.O. Box 9513, NL-2300 RA Leiden, The Netherlands}
\affiliation{European Southern Observatory, Karl-Schwarzschild-Str. 2, D-85748 Garching bei M\"unchen, Germany}
\affiliation{NASA Hubble Fellowship Program Sagan Fellow}

\author[0000-0001-7235-2417]{Kazuhiro Kanagawa} 
\affiliation{College of Science, Ibaraki University, 2-1-1 Bunkyo, Mito, Ibaraki 310-8512, Japan}

\author[0000-0001-9605-780X]{Eric W. Koch}
\affiliation{Center for Astrophysics $\mid$ Harvard \& Smithsonian, 60 Garden St., 02138 Cambridge, MA, USA}

\author[0000-0002-8896-9435]{Geoffroy Lesur} 
\affiliation{Univ. Grenoble Alpes, CNRS, IPAG, 38000 Grenoble, France}

\author[0000-0003-4663-0318]{Cristiano Longarini} 
\affiliation{Institute of Astronomy, University of Cambridge, Madingley Road, CB3 0HA, Cambridge, UK}
\affiliation{Dipartimento di Fisica, Universit\`a degli Studi di Milano, Via Celoria 16, 20133 Milano, Italy}

\author[0000-0002-8932-1219]{Ryan A. Loomis}
\affiliation{National Radio Astronomy Observatory, 520 Edgemont Rd., Charlottesville, VA 22903, USA}

\author[0000-0003-4039-8933]{Ryuta Orihara} 
\affiliation{College of Science, Ibaraki University, 2-1-1 Bunkyo, Mito, Ibaraki 310-8512, Japan}

\author[0000-0001-5907-5179]{Christophe Pinte}
\affiliation{Univ. Grenoble Alpes, CNRS, IPAG, 38000 Grenoble, France}
\affiliation{School of Physics and Astronomy, Monash University, Clayton VIC 3800, Australia}

\author[0000-0002-4716-4235]{Daniel J. Price} 
\affiliation{School of Physics and Astronomy, Monash University, Clayton VIC 3800, Australia}

\author[0000-0003-4853-5736]{Giovanni Rosotti} 
\affiliation{Dipartimento di Fisica, Universit\`a degli Studi di Milano, Via Celoria 16, 20133 Milano, Italy}

\author[0000-0002-0491-143X]{Jochen Stadler} 
\affiliation{Universit\'{e} C\^{o}te d'Azur, Observatoire de la C\^{o}te d'Azur, CNRS, Laboratoire Lagrange, France}

\author[0000-0002-3468-9577]{Gaylor Wafflard-Fernandez} 
\affiliation{Univ. Grenoble Alpes, CNRS, IPAG, 38000 Grenoble, France}

\author[0000-0002-7501-9801]{Andrew J. Winter}
\affiliation{Universit\'{e} C\^{o}te d'Azur, Observatoire de la C\^{o}te d'Azur, CNRS, Laboratoire Lagrange, France}
\affiliation{Max-Planck Institute for Astronomy (MPIA), Königstuhl 17, 69117 Heidelberg, Germany}

\author[0000-0002-7212-2416]{Lisa W\"olfer} 
\affiliation{Department of Earth, Atmospheric, and Planetary Sciences, Massachusetts Institute of Technology, Cambridge, MA 02139, USA}

\author[0000-0003-1412-893X]{Hsi-Wei Yen} 
\affiliation{Academia Sinica Institute of Astronomy \& Astrophysics, 11F of Astronomy-Mathematics Building, AS/NTU, No.1, Sec. 4, Roosevelt Rd, Taipei 10617, Taiwan}

\author[0000-0001-8002-8473	]{Tomohiro C. Yoshida} 
\affiliation{National Astronomical Observatory of Japan, 2-21-1 Osawa, Mitaka, Tokyo 181-8588, Japan}
\affiliation{Department of Astronomical Science, The Graduate University for Advanced Studies, SOKENDAI, 2-21-1 Osawa, Mitaka, Tokyo 181-8588, Japan}

\author[0000-0001-9319-1296	]{Brianna Zawadzki} 
\affiliation{Department of Astronomy, Van Vleck Observatory, Wesleyan University, 96 Foss Hill Drive, Middletown, CT 06459, USA}
\affiliation{Department of Astronomy \& Astrophysics, 525 Davey Laboratory, The Pennsylvania State University, University Park, PA 16802, USA}

\begin{abstract}

Turbulent gas motions drive planet formation and protoplanetary disk evolution. However, empirical constraints on turbulence are scarce, halting our understanding of its nature. 
Resolving signatures of the large-scale perturbations driven by disk instabilities may reveal clues on the origin of turbulence in the outer regions of planet-forming disks.
We aim to predict the observational signatures of such large-scale flows, as they would appear in high-resolution Atacama Large Millimeter/submillimeter Array observations of CO rotational lines, such as those conducted by the exoALMA Large Program.
Post-processing 3D numerical simulations, we explored the observational signatures produced by three candidate (magneto-)hydrodynamical instabilities to operate in the outer regions of protoplanetary disks: the vertical shear instability (VSI), the magneto-rotational instability (MRI), and the gravitational instability (GI). We found that exoALMA-quality observations should capture signatures of the large-scale motions induced by these instabilities. Mainly, flows with ring, arc, and spiral morphologies are apparent in the residuals of synthetic velocity centroid maps.
A qualitative comparison between our predictions and the perturbations recovered from exoALMA data suggests the presence of two laminar disks and a scarcity of ring- and arc-like VSI signatures within the sample. Spiral features produced by the MRI or the GI are still plausible in explaining observed disk perturbations. Supporting these scenarios requires further methodically comparing the predicted perturbations and the observed disks' complex dynamic structure.

\end{abstract}



\section{Introduction} \label{sec:intro}
Turbulence is a ubiquitous phenomenon in astrophysical fluids \citep{Brandenburg2011}. In protoplanetary disks, gas turbulence can transport angular momentum radially, which, together with angular momentum extraction by disk winds, is needed to explain the accretion of matter onto young stars \citep{Turner2014}. In the planet formation process, turbulence can play a major role by inducing mixing/diffusion and regulating the development of substructures \citep{Lesur2023}, which directly affects the growth and evolution of dust particles \citep{Birnstiel2024}, and the chemical inventory of the disk within planet-forming regions. Therefore, turbulence is critical in the initial stages of planet formation \citep{Drazkowska2023, Miotello2023}, and will potentially impact the composition of the forming planets. However, the dominant mechanism driving turbulence in protoplanetary disks and its role in disk evolution and planet formation are still uncertain and under active debate.

Turbulent motions span a wide range of scales. In a simplified picture, the large scales inject energy into the system, which cascades down to smaller turbulent eddies \citep[see, e.g.,][]{Lyra2019}. In protoplanetary disks, turbulent length scales larger than or comparable to the disk's local pressure scale height (roughly 10 au at 100 au from the star) could be resolvable with Atacama Large Millimeter/submillimeter Array (ALMA) observations of molecular lines. Conversely, smaller eddies within the micro scales are unresolvable with ALMA, and can take the form of line broadening \citep{Simon2015}. Contemporary methods to directly constrain turbulence levels in the disk's outer region have focused on detecting turbulent line broadening in ALMA observations of molecular lines. These measurements of `non-thermal' line broadening have reported evidence of non-zero subsonic turbulence in two disks \citep[][]{Guilloteau2012, Flaherty2020, Paneque2024, Flaherty2024}, and have set upper limits indicating relatively weak turbulence in four other disks \citep[][]{Flaherty2015, Teague2016, Flaherty2017, Flaherty2018, Flaherty2020}. However, these methods are model-dependent, subject to observational limitations and degeneracies, requiring stringent constraints on the disk structure \citep[e.g.,][Hardiman et al. in prep.]{Hughes2011, Flaherty2020,  Rosotti2023}. Further, these constraints provide limited information about the nature of turbulence in the probed outer disk regions.

High spatial resolution ALMA observations of molecular lines have revealed velocity perturbations in protoplanetary disks \citep{DiskDynamics2020, Pinte2023}. Such perturbations, observed as deviations from the background disk rotation, have opened the possibility to hunt for signatures of large-scale motions driven by turbulence mechanisms \citep{Hall2020, Barraza2021}. Identifying signatures from disk instabilities has the potential to unravel the origin of turbulence in the disk's outer regions. Moreover, since disk instabilities require specific conditions to develop and be sustained, their detection can constrain the disk's physical properties. Nevertheless, finding evidence of large-scale turbulent motions in disks has been challenging, as it requires high spectral and spatial resolution. Such quality has been delivered by the exoALMA survey \citep{Teague_exoALMA}, providing a detailed view of the disk kinematic structure of 15 protoplanetary disks. These new observations allow us to hunt for signposts of turbulence in the large-scale flow predicted by 3D numerical simulations of disk instabilities.

Various instabilities are expected to be active and to drive turbulence in the disk outer regions \citep{Lesur2023}. Gravitational instabilities (GI) are predicted to develop in massive protoplanetary disks \citep{Toomre1964, Kratter2016}. The magneto-rotational instability (MRI) can operate in ionized regions of magnetized protoplanetary disks \citep{Balbus1991, Hawley1991}, while purely hydrodynamical instabilities can dominate in the weakly ionized regions \citep{Klahr2018, Lyra2019}. Among the hydro instabilities, the vertical shear instability (VSI) can operate in fast-cooling regions of the outer disk \citep{Urpin1998, Nelson2013, Pfeil2019, Cui2020, Cui2022}, while the Convective Overstability \citep[COV,][]{Klahr2014,Lyra2014} and Zombie Vortex Instability \citep[ZVI,][]{Marcus2015,Marcus2016} are unlikely to be dominant in the bulk of the disk outward tens of au from the star \citep{Malygin2017,Pfeil2019}.
These instabilities give rise to disk substructures that are distinct from each other and from other substructure-forming processes, such as planet-disk interactions \citep{Bae2023}. Therefore, large-scale turbulent motions detected by molecular line emission of protoplanetary disks may unravel the instability at play from its characteristic observational features.

This work explores the kinematic signatures of large-scale gas motions produced by instability candidates to drive turbulence in the outer regions of planet-forming disks: the VSI, MRI, and GI. We present our results in the context of the exoALMA Large Program, addressing its potential to unravel large-scale turbulence signatures. The paper is organized as follows: We describe the methods for the numerical simulations in Section \ref{simulations}, and we detail the methods used for the radiative transfer predictions, mock observations, and calculation of observables in Section \ref{postprocessing}. We show the simulations' results in Section \ref{simulationsresults}, and their post-processing results in Section \ref{rtresults}. We discuss the findings and caveats of our work in Section \ref{discussion}, and we draw our conclusions in Section \ref{conclusions}.


\section{Methods for Numerical Simulations}\label{simulations}

The numerical simulations presented in this work were conducted with upgraded versions of the grid-based code \textsc{PLUTO}\footnote{\url{https://plutocode.ph.unito.it/}} \citep[][see details of upgrades in \citealt{Flock2015, Flock2020, Bethune2021}]{Mignone2007}. We post-processed outputs from three 3D global (magneto-)hydrodynamic simulations, one representative of each of the disk instabilities explored. Particularly, we analyzed an individual simulation snapshot for each of the studied instabilities taken after the disk had reached a quasi-steady state, which we considered representative of their gas dynamics.

\subsection{VSI-unstable disk simulation}
The disk gas dynamics induced by the VSI is explored with outputs of the 3D global hydrodynamic simulation with radiative energy transport (henceforth RHD) presented in \citet{Flock2020}.
The simulation was conducted with PLUTO version 4.2, using the setup of the numerical simulations performed in \citet{Flock2017}. The radiation hydrodynamic equations are solved using a hybrid stellar irradiation and flux-limited diffusion method \citep[][see further details in \citealt{Flock2020} Section 2]{Flock2013}. The parameters were chosen to simulate a disk around a typical T Tauri star with a stellar mass of $M_{\star}=0.5 M_{\odot}$.
The simulation domain covers 20-100 au in the radial domain, 0.7 rad in colatitude ($\pi/2 \pm 0.35 \, \rm rad$, with the midplane at $\theta = \pi/2$) and $2 \pi$ rad in azimuth. The simulation of the nonlinear evolution of the VSI was run with a grid with $1024$, $512$, and $2044$ cells in $r$, $\theta$ and $\phi$, respectively, with a mesh logarithmically spaced in the radial direction and uniformly spaced in colatitude and azimuth.
However, in order to reduce the cost of the radiative transfer calculations, we post-processed outputs interpolated to a coarser grid, half the grid resolution of the original in each direction, that is, $512$, $256$, and $1022$ cells in $r$, $\theta$, and $\phi$, respectively. The domain in colatitude covers approximately $7.3$ pressure scale heights, with approximately 70 cells per pressure scale height.
Buffer zones were applied in the innermost $2$ au and outermost $3$ au of the radial domain.
For the boundary conditions, a modified outflow condition was used in the $r$ and $\theta$ directions. The condition prevents gas inflow at the domain boundaries. Additionally, at the boundaries in colatitude the gas density was extrapolated logarithmically into the ghost zones \citep[see Section 2 in][]{Flock2020}.
For a summary of the simulation parameters, see Table \ref{tab:simulationsparameters} (see also Table 1 in \citealt{Flock2020}).

\subsection{MRI-unstable disk simulation}

The post-processing of the gas dynamics of a MRI turbulent protoplanetary disk is performed with outputs from the magneto-hydrodynamic simulation (henceforth MHD) presented in \citet{Flock2015}. In particular, the 3D global non-ideal MHD stratified simulation labeled \texttt{D2G\_e-2}, ran for a disk with a dust-to-gas mass ratio of $10^{-2}$. Only the 2D ($R,\theta$) disk temperature structure was obtained using radiative transfer calculations, and it remained fixed via a prescribed locally isothermal equation of state \citep[details in Section 2 of][]{Flock2015}.
The non-ideal MHD equations were solved using the MHD module of PLUTO \citep{Mignone2007} including Ohmic diffusion. Our simulation neglects the Hall effect and ambipolar diffusion. Similarly to the Ohmic diffusion, ambipolar diffusion is a dissipative effect and can further damp the MRI, especially in the disk's outer regions \citep{Bai2015,Simon2018}, and it can drive additional annular substructures \citep{Kunz2013, Simon2014, Cui2021}. Distinctively, the Hall effect plays a role in transporting magnetic flux in the disk \citep{BaiStone2017}, but is not dissipative and can lead to higher magnetic stress, driving accretion \citep{Lesur2014, Bethune2017, BaiStone2017}. Overall, the Hall effect typically stabilizes small scales against the MRI \citep[e.g.,][]{BalbusTerquem2001} and also facilitates large-scale structure formation \citep{Kunz2013, Bethune2016, Krapp2018}. Therefore, the inclusion of ambipolar diffusion and the Hall effect will modify the morphology of disk flow structures increasing the complexity of the disk kinematic structure, which can difficult the observability of MRI-turbulence via gas kinematics. Moreover, ambipolar diffusion can also weaken the MRI-induced velocity perturbations in the outer disk regions, which is not optimal for exploring the observability of MRI turbulence via ALMA CO kinematics. Nevertheless, the Hall effect is predicted to be relevant in the midplane region of the outer disk \citep{Bai2015}, which is challenging to probe with ALMA molecular line observations.

Similar to the RHD simulation, the MHD simulation parameters are set to simulate a typical T Tauri system, with a central star of $M_{\star}=0.5 M_{\odot}$ and a disk mass of $0.085 M_{\star}$.
The radial and azimuthal domain of the MHD simulation are equal to the set for the RHD simulation, while $0.72$ rad in colatitude. In this case, the domain in $\theta$ covers approximately $6.4$ disk pressure scale heights, $3.2H$ at each disk hemisphere.. The MHD simulation grid was set to $256\times 128\times 512$, in $r$, $\theta$, and $\phi$, respectively ($\approx 20$ cells per $H$). 
The boundary condition in radius and colatitude continue the gradient of the gas density, pressure, azimuthal velocity and resistivity. For the velocities in the $r$ and $\theta$ directions, a zero gradient condition was used, with a linear damping of the normal component the case of gas moving toward the simulation domain. In addition, the gas density is prevented from increasing in the $\theta$ ghost cells \citep[see Section 2 in][]{Flock2015}.
Further details on the MHD simulation's parameters are presented in Table \ref{tab:simulationsparameters} of the Appendix \citep[see also Table 1 in][]{Flock2015}.

\subsection{GI-unstable disk simulation}
To explore the case of a disk with active gravitational instability, we post-processed an output from a self-gravitating hydrodynamic simulation (henceforth SGHD) presented in \citet{Bethune2021}. In particular, we used the outputs of the model labeled \texttt{M5B10} in \citet{Bethune2021}, which corresponds to a disk with an initial disk-to-star mass ratio of $1/5$ and a dimensionless constant cooling parameter $\beta$ of $10$, setting the time-scale for radiative cooling ($t_{\textrm{cool}}=\beta \Omega_{\star}^{-1}$, with $\Omega_{\star}$ the local Keplerian frequency; see Equation 5 in \citealt{Bethune2021}). 
The radial domain of the SGHD simulation extends from 5 to 160 au (1 to 32 code units of length, set to 5 au), covered with 518 cells following a logarithmic spacing. 
The domain in $\phi$ covers the full $2\pi$ rad in the azimuthal direction with uniformly sampled $512$ cells. In colatitude, the domain is covered by $96$ cells in total, extending from $\theta = \pi/2 -0.35$ rad to $\theta = \pi/2 +0.35$ rad, where the grid has a finer sampling in the $0.1$ rad covering the midplane region ($\pi/2 \pm 0.05 \rm\, rad$), meshed with 32 uniformly spaced cells. The regions at the disk's upper layers are covered by a coarser mesh, with grid spacings stretched in $\theta$ \citep[see Section 2.5.1 in][]{Bethune2021}. Near the midplane the simulation has approximately $9$ cells per $H$.
For the simulation boundaries in colatitude ($\theta$), symmetric and antisymmetric boundary conditions are set for the gas density and velocities, respectively. In the radial direction, a symmetric boundary condition is set for $\rho$ and $v_{\phi}/r$, while an antisymmetric condition is used for $v_{\theta}$ and $r^2 v_r$. In addition, a small pressure is imposed in the ghost cells for all boundaries \citep[see Section 2.5.4. in][]{Bethune2021}.
The parameters used for the SGHD simulation are presented in Table \ref{tab:simulationsparameters} \citep[see also Table 1 in][]{Bethune2021}.

\subsection{Considerations for post-processing}

Turbulence levels are expected to be strongest at the uppermost layers in disks with active VSI or MRI \citep[see, e.g.,][]{Simon2015, Flock2015, Pfeil2021}. Therefore, we focus our analysis and discussion on the surface disk layers, which are the regions probed by $\rm ^{12}CO(3-2)$ line emission observed by exoALMA (within ALMA Band 7), with typical $Z/R$ values of $\sim 0.27$ \citep{Galloway_exoALMA}. Deeper layers of the disk (closer to the disk midplane), as traced by $\rm ^{13}CO(3-2)$ and $\rm CS(7-6)$ exoALMA observations (with typical $Z/R$ values of $\sim 0.17$ and $\sim 0.18$, respectively \citealt{Galloway_exoALMA}), will trace regions of weaker turbulence, in particular for the VSI and the MRI cases \citep{Flock2011, Simon2018, Pfeil2021}. In addition, the signal-to-noise ratios (SNRs) from the $\rm CS(7-6)$ line observations are not sufficient to fully exploit the spectral resolution achieved by exoALMA \citep{Teague_exoALMA}, crucial to capture the subtle gas motions driven by disk instabilities.

Due to the inclusion of radiation-hydrodynamics in the RHD simulation \citep{Flock2020}, and the use of a disk temperature obtained from radiative transfer calculations in the MHD simulation \citep{Flock2020}, it is not straightforward to adapt the code unit of length to re-scale the simulation domain to model a larger disk, more suited to compare with the exoALMA sources. In particular, the RHD and MHD simulations' disks temperature structures are set to model the radial range between $20$ au and $100$ au, while the majority of exoALMA disks are significantly larger than $100$ au in size. Since the size of disk substructures in the gas increases with disk radius \citep[see, e.g.,][]{Bae2023}, proportional to the disk pressure scale height, structures at larger radius will be easier to resolve spatially. However, if our RHD and MHD simulations are re-scaled to larger disk sizes, the size of substructures in the synthetic predictions would be underestimated. Therefore, to account for the difference in disk sizes between our models and the exoALMA disks when exploring the observability of kinematic structures (Section \ref{convolvedmodels}), we convolved the synthetic images by a smaller synthesized beam than that of the fiducial exoALMA images (see  Section \ref{methodscolvolved}). Our convolved images have a spatial resolution such that the substructures we resolve in our predictions can be compared to the ones we observe in the outer regions of the exoALMA sources, allowing a fair comparison between synthetic observations and exoALMA data (see Section \ref{convolvedmodels}).
Finally, to compare the same radial region between the three simulations, we only explore the region between $20$ au and $100$ au of the GI simulation, truncating the innermost $15$ au and outermost $60$ au in the post-processing.

\begin{figure*}[htp]
\centering
\includegraphics[angle=0,width=1.0\linewidth]{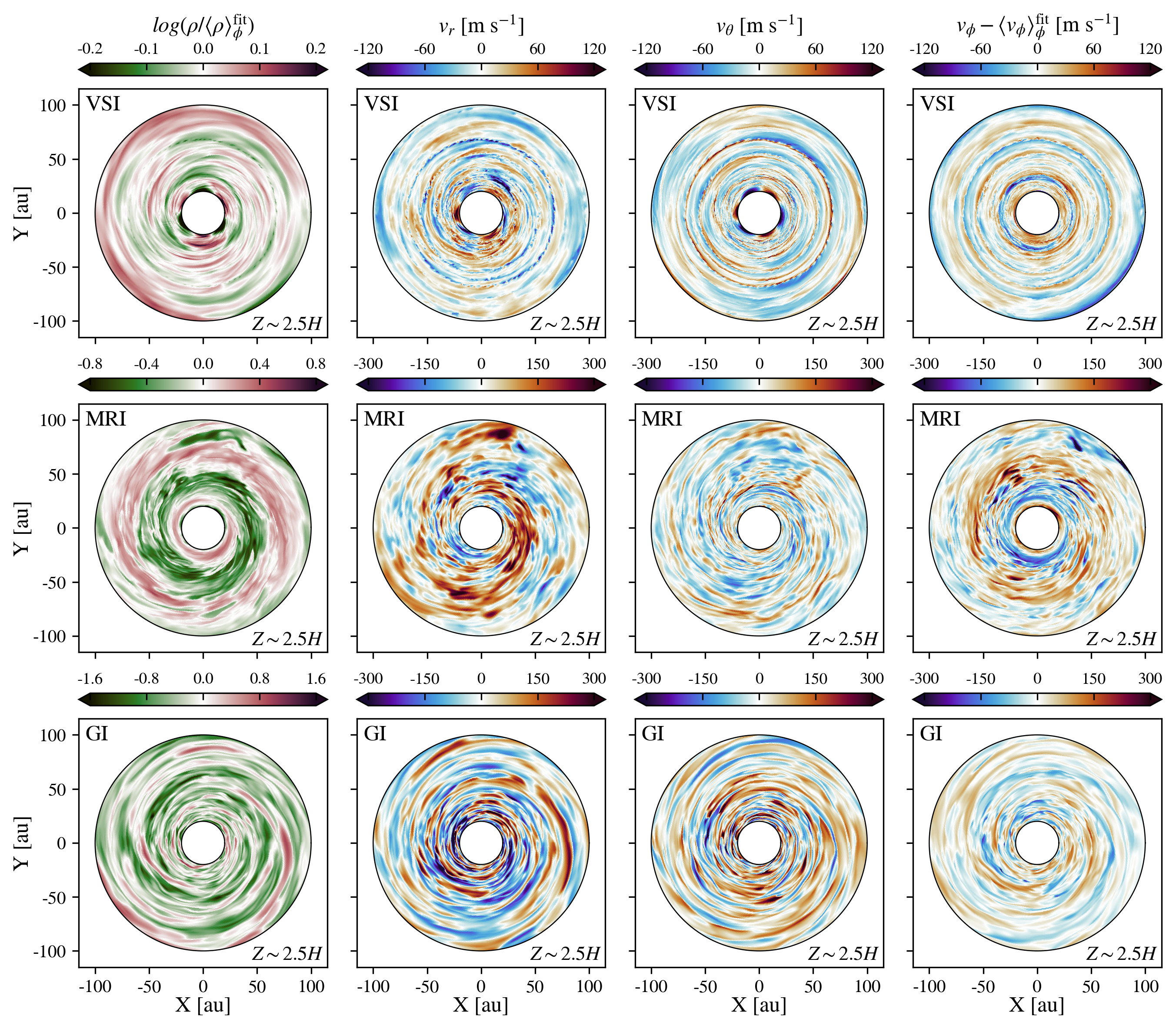}
\caption{Cartesian view of the disk perturbations at $Z\sim 2.5H$ for the HD (top), MHD (middle), and SGHD (bottom) simulations, with active VSI, MRI, and GI, respectively. From left to right, are shown perturbations in the gas density, radial velocity, meridional velocity and azimuthal velocity. The color bar limits are adjusted for each instability.}
\label{fig:simulationssurface}
\end{figure*}


\section{Methods for Radiative Transfer Models}\label{postprocessing}

We post-processed the numerical simulations' outputs with the Monte-Carlo radiative transfer code \textsc{RADMC-3D}\footnote{\url{https://www.ita.uni-heidelberg.de/~dullemond/software/radmc-3d/}} \citep{Dullemond2012} version 2.0. We build our \textsc{RADMC-3D} input files using scripts based in \textsc{fargo2radmc3d}\footnote{\url{https://github.com/charango/fargo2radmc3d}} \citep{Baruteau2019, Baruteau2021}, but built for outputs from \textsc{PLUTO} \citep[see further details in][]{Barraza2021,Barraza2024}. The adopted framework is analogous to the exoALMA benchmark, presented and validated in \cite{Bae_exoALMA}.

\subsection{Radiative Transfer Setup}\label{rtsetup}

Our disk model considered only one hemisphere of the disk (the front half of the disk as oriented in the sky, also referred to as the upper part of the disk; for a schematic view, see Figure 2 in \citealt{Pinte2019}). We set the number density of $\rm ^{12}CO$ to zero for the bottom half of the disk, having only the contribution from the upper disk $\rm ^{12}CO$ layer into the line emission in our predictions. This choice allows a more direct comparison with model residuals obtained from observations, which, with a complex line fitting, separate the line emission contribution from the upper and bottom sides \citep[see, e.g.,][]{Casassus2021, Izquierdo2022, Izquierdo_exoALMA}. Since we are only interested in studying the perturbations obtained from the upper layer of the disk, not including the bottom side in our models avoids complications in the analysis that would come from the mixing of line emission from both upper and bottom $\rm ^{12}CO$ layers.
We adopted a constant value throughout the upper half of the disk of $10^{-4}$ for the fraction of $\rm ^{12}CO$ relative to H$_2$ ($n(\rm ^{12}CO$)/$n(\rm H_2)$), including a highly simplified CO photo-dissociation and freeze-out prescriptions. Relevant toward the midplane region of the disk, CO freeze-out was included by reducing the CO number density by a factor $10^{-5}$ in regions where the gas temperature falls below $21$ K, assuming that CO is then frozen onto dust grains \citep[e.g.,][]{Schwarz2016, Bae2021}. Important at the disk surface, photo-dissociation of $\rm ^{12}CO$ was implemented by reducing its number density $n$($\rm ^{12}CO$) by a factor of $10^{-5}$ above the disk heights where the vertically-integrated gas column density, computed from the surface towards the midplane, has a value lower than $10^{21}$ cm$^{-2}$ \citep[e.g.,][]{Visser2009, Simon2015, Flaherty2015, Bae2021, Baruteau2021}. The freeze-out prescription allows to recover a geometrically thin layer of $\rm ^{12}CO(3-2)$ emission, while photo-dissociation helps predicting the emission layer at a height similar to empirical constraints, as obtained from resolved $\rm ^{12}CO(3-2)$ observations of protoplanetary disks (typical $Z/R\sim 0.27$ in exoALMA disks, \citealt{Galloway_exoALMA}; see also, e.g., \citealt{Law2023, Paneque2023}). In addition, including photo-dissociation avoids possible artifacts from the edge of the simulation domain in colatitude in the line ray-tracing.

For simplicity and to have the same temperature structure for all disk models, we assumed a parametric temperature profile that takes into account radial and vertical gradients (\citealt {Flaherty2020}; see also, e.g., \citealt{Dartois2003, Rosenfeld2013a, Flaherty2015,Galloway_exoALMA}):

\begin{equation}
    T_{\rm mid} = T_{\rm mid0} \left( \frac{R}{150 \, \rm au} \right) ^q,
\end{equation}
\begin{equation}
    T_{\rm atm} = T_{\rm atm0} \left( \frac{R}{150 \, \rm au} \right) ^q,
\end{equation}
\begin{equation}
    T_{\rm gas} =
    \begin{cases}
    T_{\rm atm}+(T_{\rm mid}-T_{\rm atm}) \left[ \cos{\frac{\pi Z}{2 Z_{\rm q}}} \right]^2, & \text{if } Z< Z_{\rm q} \\
    T_{\rm atm},              & \text{otherwise}
    \end{cases}
\end{equation}
\begin{equation}
    Z_{\rm q} = Z_{\rm q0} \left( \frac{R}{150 \, \rm au} \right) ^{1.3},
\end{equation}
where $R$ and $Z$ are the disk cylindrical disk coordinates, corresponding to the disk radius and vertical height, respectively. $Z_{\rm q}$ denotes the height from the disk midplane where the gas temperature reaches its maximum, set by $T_{\rm atm}$. We assumed the parameters of the fiducial model for DM Tau presented in \citealt{Flaherty2020}, that is, $T_{\rm atm0} = 24.68$ K, $T_{\rm mid0} = 14.3$ K, $q=-0.371$, and $Z_{\rm q0}=70$ au \citep[see also][]{Galloway_exoALMA}. However, the selection of our model temperature structure and parameters is not to reproduce a specific target, but to consider a disk structure that follows empirical constraints from ALMA observations of molecular lines. Two-dimensional views of our disk models' gas temperature and $\rm ^{12}CO$ number densities are shown in Figure \ref{fig:numberdensmodels} of the Appendix.

For the line ray-tracing of the synthetic images, we assumed a distance to the source of $140$ pc \citep[approximately the average distance of the nearest $13$ exoALMA sources,][]{Teague_exoALMA}. We assumed a disk position angle $\rm PA$ of $90^{\circ}$ east of north, where the south part of the disk is nearest to the observer, and the disk rotation is counterclockwise. We computed images for five disk inclinations ranging from $i=5^{\circ}$ to $i=60^{\circ}$, in order to cover the range of inclinations of the exoALMA disks \citep{Teague_exoALMA}. Finally, to predict the \emph{raw} (unconvolved) kinematic signatures from large-scale turbulence (presented in Section \ref{rawkinematicsignatures}), we considered the case of achieving ALMA's highest spectral resolution available in Band 7 at the frequency of the $\rm ^{12}CO$ $J:3-2$ rotational transition ($\approx 345.796 \rm\, GHz$). Therefore, we produced data cubes with a $26 \rm \,m\,s^{-1}$ resolution (shown in the left $2.5$ panels of Figure \ref{fig:chanelmaps}, and in Figure \ref{fig:nonkeplerian}).

\subsection{Synthetic exoALMA observations}\label{methodscolvolved}

For the case of predictions at exoALMA resolution (presented in Section \ref{convolvedmodels}), we binned the raw data cubes in frequency down to a velocity resolution of $104 \rm \,m\,s^{-1}$ using CASA\footnote{\url{https://casa.nrao.edu/}} \texttt{imrebin} task \citep{CASA}, matching approximately the velocity resolution of the fiducial exoALMA data cubes for $\rm ^{12}CO(3-2)$ \citep{Teague_exoALMA, Loomis_exoALMA}.

We explored the effect of the spatial resolution of ALMA observations by convolving our raw radiative transfer synthetic images with a circular Gaussian beam. Due to the difference in the disk sizes between our models and the observed disk sizes, we consider a Gaussian beam with a FWHM of $35$ mas, while the exoALMA fiducial images have a $\sim 150$ mas spatial resolution \citep{Teague_exoALMA}. Our disk models have a radial extent of $100$ au, while the average value of the flux based disk radius enclosing the $95\%$ of $\rm ^{12}CO(3-2)$ emission, $R^{95}_{\rm 12CO}$, is $411$ au for the exoALMA sample \citep{Galloway_exoALMA}. Therefore, with our predictions' spatial resolution, the synthetic observations maintain a similar number of independent beams across the disk as the fiducial exoALMA images for the average exoALMA disk size (roughly $\sim 40$ beams across the disk diameter). This results in a comparable amount of resolved disk substructures to those present in the exoALMA observations.
Before convolving the images with the Gaussian beam, we include noise in our models as white noise, resulting in an RMS of $\sim 0.16$ mJy beam $^{-1}$ after convolution. The target RMS value is set to be equivalent to the achieved by the fiducial images of the exoALMA survey $\rm ^{12}CO(3-2)$ observations \citep[$\sim 3$ mJy beam $^{-1}$ for a larger beam of 150 mas, and $100$ m s$^{-1}$ velocity resolution,][]{Teague_exoALMA}.

\subsection{Line Analysis Tools}\label{kinematictools}

To compute the observables from the raw and convolved synthetic images we used the Python package \textsc{bettermoments}\footnote{\url{https://github.com/richteague/bettermoments}} \citep{bettermoments}. We applied the percentiles method from \textsc{bettermoments} to collapse each data cube and obtain the line centroid and line width at every image pixel. The percentile method collapses the cube by taking the percentiles of the intensity-weighted velocity, for which the median returns the line centroid, and the half of the range between the 16th and 84th percentile returns an accurate line width measurement. To obtain the $^{12}$CO temperature, we computed the line peak maximum at each image pixel.

We performed the analysis of the observables obtained from the synthetic convolved images using the Python packages \textsc{eddy}\footnote{\url{https://github.com/richteague/eddy}} \citep[Extracting Disk Dynamics][]{eddy} and \textsc{GoFish}\footnote{\url{https://github.com/richteague/gofish}} \citep{GoFish}. 
We used \textsc{eddy} to obtain the best fit modified Keplerian disk model for an elevated surface to our synthetic $\rm ^{12}CO(3-2)$ line centroid maps. The modified Keplerian disk rotation profile includes an additional radial-dependent component that can account for super-Keplerian or sub-Keplerian deviations from a purely Keplerian disk \citep[see details in][also included in Section \ref{app:diskmodeleddy} of the Appendix]{Teague2022}. The flexible rotation profile helps to capture the contributions from the gravitational potential of the disk and the pressure support in the disk's outer edge \citep[see Figure 14 in][]{Teague2022}, giving an improved description of the background bulk rotation of the disk. The modified Keplerian model is particularly important in properly describing the super-Keplerian background rotation of the self-gravitating disk simulation.

To obtain our best fit disk models, we computed the posterior distribution of our free parameters with \texttt{eddy} (see details in Section \ref{app:diskmodeleddy} of the Appendix). By subtracting the best-fit model of the projected disk bulk rotation from the perturbed disk line centroid map, the observable kinematic substructures driven by the disk instabilities can be explored in the residuals (presented in Section \ref{rtresults}).

To obtain the line width and temperature maps residuals, we used \textsc{GoFish} to subtract the disk azimuthal average in the $\rm ^{12}CO(3-2)$ line peak (temperature), and line width maps, which takes into account the elevated emission surface used as input into the \textsc{eddy} fitting. These residual maps retrieve the non-axisymmetric substructures present in the disk, induced by deviations in the traced temperature and line broadening relative to the disk background.


\section{Results from Numerical Simulations}\label{simulationsresults} 

In the following, we present the outputs of the set of numerical simulations of turbulent protoplanetary disks post-processed with radiative transfer calculations (see Section \ref{postprocessing}). We highlight the distinct morphology of the perturbations driven by the VSI, MRI, and GI in the gas density and velocities.

Figure \ref{fig:simulationssurface} shows the perturbations driven by the VSI, MRI, and GI at $2.5$ pressure scale heights ($H$) from the disk midplane. The layer shown is chosen as a representative snapshot of the gas dynamics at the upper layers, as traced by $\rm ^{12}CO(3-2)$. We present the normalized gas density relative to a radial power-law fit to the azimuthal average in logarithmic scale ($log_{10}(\rho/\langle \rho \rangle_{\phi})$), gas radial velocity ($v_r$), meridional (or vertical) velocity ($v_{\theta}$), and azimuthal velocity deviations relative to a radial power-law fit to its azimuthal average ($v_{\phi}-\langle v_{\phi}\rangle_{\phi}^{\textrm{fit}}$). The fitting of a radial power-law to the azimuthal average of the azimuthal velocity, for which we use \texttt{scipy.optimize.curve\_fit}, allows a superb subtraction of the disk bulk rotation, extracting the perturbations in $v_{\phi}$. We also show the perturbations at the disk midplane in Figure \ref{fig:simulationsmidplane} of the Appendix for completeness.

In the VSI snapshots (first row of Figure \ref{fig:simulationssurface}), we display the characteristic dynamics from the non-linear state of a VSI-unstable disk, in which the instability has reached a saturated state. In the gas density, VSI drives subtle perturbations, where the maximum density contrast within annuli at $Z=2.5H$ has a median value of $\approx 15\%$ (computed within $35$ and $85$ au). Such weak density perturbations are likely not relevant in observations of line emission of protoplanetary disks. However, in the gas velocity the VSI drives substantial perturbations in the three velocity components. In particular,  the VSI induces a characteristic corrugated flow pattern in the meridional direction, which displays a high degree of axisymmetry \citep[see, e.g.,][]{Flores2020,Barraza2021}. These velocity perturbations have an average RMS velocity at $Z\sim 2.5H$ of $\sim 1\%$ the local Keplerian velocity. In CO kinematic observations, the apparent difference in axisymmetry of the velocity perturbations among velocity components becomes relevant when exploring their projection into the line-of-sight, in which inclined disks will have a larger contribution from the radial and azimuthal velocities (see Section \ref{rtresults}).

For the MRI simulation (middle row of Figure \ref{fig:simulationssurface}), the snapshots also reveal that the MRI-driven gas density structure is dominated by a radial gap and overdensity induced by the presence of an outer MRI dead-zone edge, a radial transition in the disk turbulence levels \citep{Flock2017}. A steep transition in turbulence levels develops in the simulation owing to the inclusion of Ohmic diffusion. However, the location of the MRI dead-zone outer edge and its dynamics can be substantially affected if the Hall effect and ambipolar diffusion are also taken into account in the non-ideal MHD simulation \citep[e.g.,][see also Section \ref{simulations} and \ref{caveats}]{Lesur2014}. In particular, the Hall effect could `revive' regions of the dead-zone \citep{Kunz2013, Lesur2014}.

Azimuthally-asymmetric perturbations in the gas density at $Z\sim 2.5H$ remain below $65\%$ relative to the background. In the velocities, the perturbations driven by the MRI are larger than VSI-driven ones, with an average RMS velocity at $Z\sim 2.5H$ of $\sim 3.6\%$ the local Keplerian velocity, roughly $3.6$ times larger than the velocity perturbations driven by the VSI. Distinctively, MRI-induced perturbations show radial flows that are substantially stronger than those in the meridional and azimuthal directions, where the latter two have similar magnitudes. At $Z\sim 2.5H$, the mean RMS velocity of the radial perturbations of the MRI is approximately a factor $1.46$ larger than the mean RMS velocity of the ones in the meridional and azimuthal directions. In addition, its morphology shows a spiral structure, composed of multiple spirals arms (high modes spirals are usually dominant for the MRI, see Section 5.1.2 in \citealt{Bae2023} and references therein). The MRI-driven spiral structure is particularly defined in the radial and meridional velocity components (see also Fig. \ref{fig:simulationspolar} for a polar deprojection of the simulation fields). In the azimuthal velocity, rings of super- and sub-Keplerian velocities are present, induced at outer MRI dead-zone outer edge.
Overall, the velocity structure driven by MRI shows a turbulent-like morphology, with localized flows and `meso-scale' eddies (with scales smaller than the gas pressure scale height) distorting the large-scale spiral structures and non-Keplerian rings. These turbulent features might complicate the identification of MRI-driven large-scale coherent structures \citep[e.g., with filament detection methods;][]{Koch2016, Izquierdo2022, Izquierdo2023,Izquierdo_exoALMA}, and could lead to misidentifying localized deviations when searching for embedded planets' signposts, further discussed in Section \ref{rtresults}.

GI-driven perturbations (bottom row of Figure \ref{fig:simulationssurface}) show a characteristic large-scale spiral structure, exhaustively studied in previous works \citep[e.g.,][]{Lodato2004, Cossins2009, Forgan2011, Dong2015, Hall2018, Bethune2021}. These spiral structures are present in the three velocity components, with common pitch angles (see Figure \ref{fig:simulationspolar}). From Figure \ref{fig:simulationssurface}, we observe that the density perturbations induced by GI are substantially larger than those driven by the VSI or the MRI. Therefore, the GI-driven density perturbations have a higher chance to create substructures in the $\rm ^{12}CO(3-2)$ line peak and line width, that we address in further detail in Section \ref{rtresults}. Similarly, the velocity perturbations driven by the MRI and the GI are substantially stronger than those driven by the VSI. In our sample of simulations, the average RMS velocities between $25$ and $95$ au at $Z\sim 2.5H$ are $29 \, \rm  m\,s^{-1}$, $111 \, \rm m\,s^{-1}$ and $\rm 142\, m\,s^{-1}$, for the VSI, the MRI and the GI, respectively. Therefore, the GI is more likely to create observable features in $\rm ^{12}CO(3-2)$ kinematic observations. For completeness, we show the entire domain of the SGHD simulation in Figure \ref{fig:gisimulationfull} of the Appendix.

Differences in magnitude between the velocity perturbations at the disk midplane and upper layers for the MRI and the VSI cases are evident when comparing the Cartesian view of the perturbations at $Z\sim2.5H$, shown in Figure \ref{fig:simulationssurface}, and the perturbations at $Z\sim 0$, provided in Figure \ref{fig:simulationsmidplane} of the Appendix. However, in the case of perturbations driven by GI, the radial and azimuthal motions are still vigorous near the midplane, showing a decrease in magnitude toward the midplane for the meridional velocity component only, consistent with previous findings \citep{Shi2014}.
In our sample of simulations, the average RMS velocity moving down from $Z\sim 2.5H$ to $Z\sim 1.5H$ shows a decrease of $\sim 47 \%$ for the VSI, $\sim 62 \%$ for the MRI, while only $\sim 18 \%$ for GI. The weak dependency of the GI-induced RMS turbulent velocities with disk height aligns with previous studies \citep[e.g.,][]{Shi2014, Forgan2012}. Therefore, mapping deeper layers with molecular line observations should still reveal GI perturbations \citep[see also][]{Longarini2021b}, but less likely VSI- or MRI-driven large-scale motions.

Finally, we stress that our unstable disks simulations are for particular sets of parameters, which control the resulting strength of the instabilities. In conditions closer to the ideal MHD limit, the MRI can be stronger \citep[e.g.,][]{Flock2015}. Conversely, ambipolar diffusion can further decrease the strength of the MRI \citep[e.g.,][see also Section \ref{simulations}]{Simon2015}. In the case of GI, the strength of the perturbations tightly depends on the disk mass and its cooling properties \citep{Bethune2021}. For the VSI, a steep disk thermal stratification enhances the vertical shear, driving stronger VSI turbulence \citep{Yun2025a, Yun2025b}. Therefore, our observability study focuses on the differences in morphology of the kinematic signatures rather than their strength. Nevertheless, we comment on the variations in magnitude of the velocity perturbations and their impact on the observable signatures. Still, the explored observable kinematic features must be taken as a reference from a particular set of simulations, for which the post-processed snapshots were considered to be representative of the overall dynamics of the studied disk instabilities.


\section{Results from Radiative Transfer Models}\label{rtresults}

\begin{figure*}[htp]
\centering
\includegraphics[angle=0,width=\linewidth]{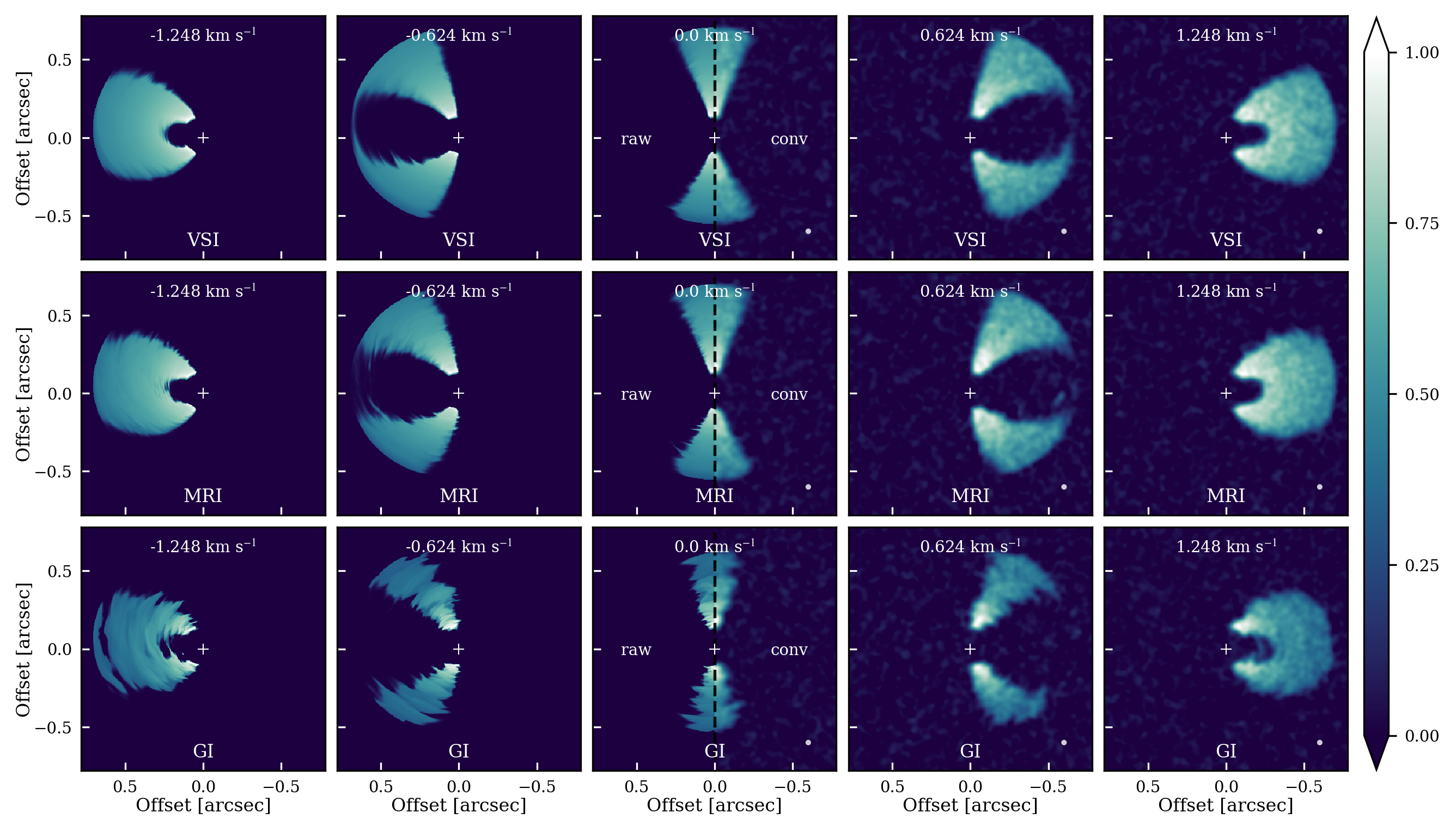}
\caption{Selected channel maps from $\rm ^{12}CO(3-2)$ predictions of disks unstable to the VSI, MRI and GI, for an disk inclination of $30^{\circ}$. The left $2.5$ panels show the raw product, while the right $2.5$ panels include the effect of the spatial resolution ($35$ mas FWHM Gaussian beam), spectral resolution ($104$ m s$^{-1}$ channel spacing), and sensitivity of exoALMA. The models only consider the upper (or front) CO layer.}
\label{fig:chanelmaps}
\end{figure*}

\begin{figure*}[htp]
\centering
\includegraphics[angle=0,width=\linewidth]{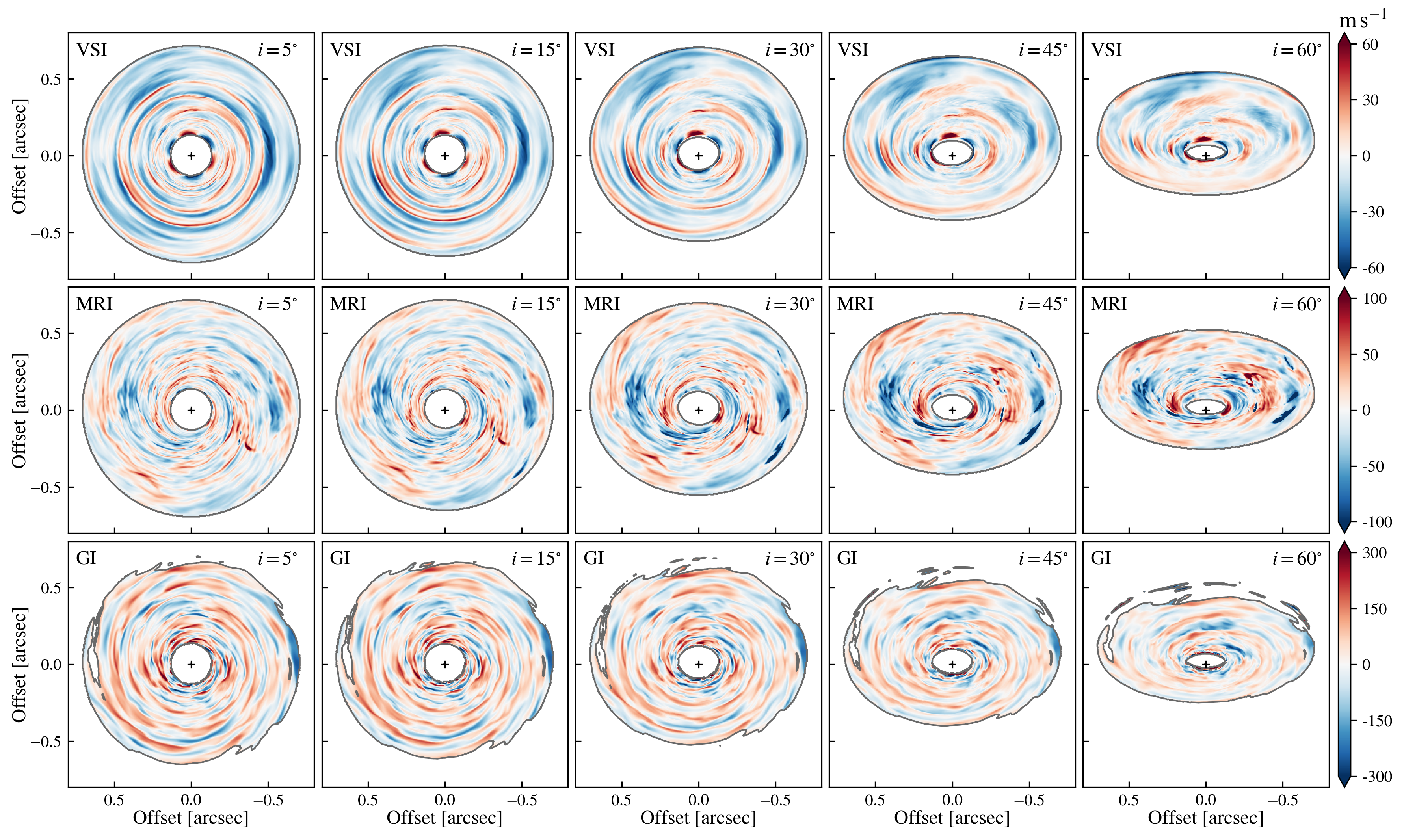}
\caption{Line centroid map residuals of the disk velocity relative to its rotation from \emph{raw} $\rm ^{12}CO(3-2)$ predictions of disks unstable to the VSI, MRI and GI, assuming a perfect background subtraction of the disk rotation. The raw images have a $26$ m s$^{-1}$ velocity resolution. From left to right, different disk inclinations are shown. The models consider only the upper (or front) disk hemisphere, and the disk rotation is counterclockwise in the sky. The color bar limits have been adjusted for each instability.}
\label{fig:nonkeplerian}
\end{figure*}

\begin{figure*}[htp]
\centering
\includegraphics[angle=0,width=\linewidth]{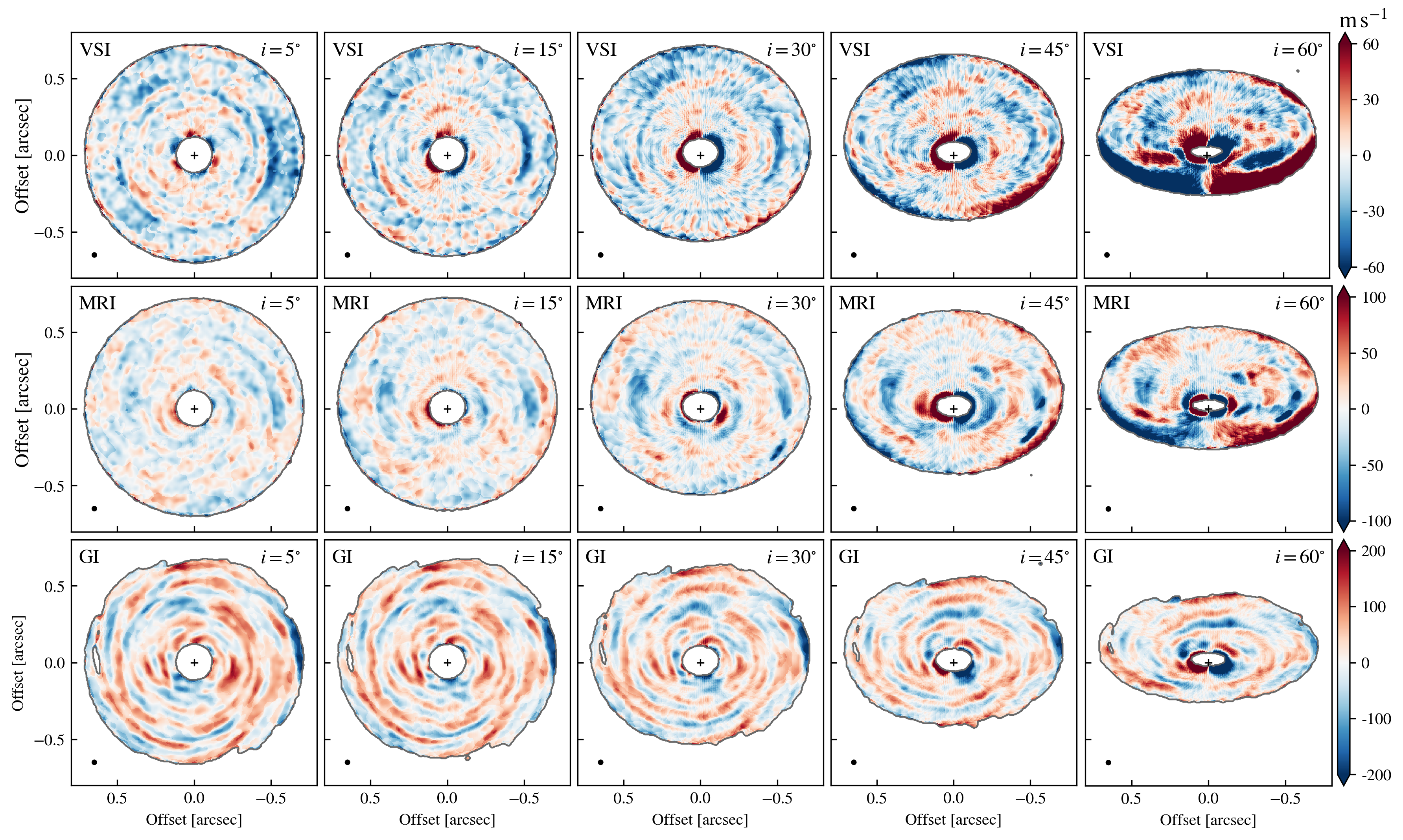}
\caption{Line centroid map residuals of the disk velocity relative to its rotation from $\rm ^{12}CO(3-2)$ predictions of turbulent protoplanetary disks, including the effects of the spatial resolution ($35$ mas FWHM Gaussian beam), the spectral resolution ($104$ m s$^{-1}$ channel spacing), and the sensitivity of exoALMA. We show the resulting velocity deviations from a modified Keplerian model for a geometrically thick disk, fitted to the line centroid map using \textsc{eddy}. The disk models consider only the upper (or front) $\rm ^{12}CO$ layer, and the disk rotation is counter clockwise in the sky. From top to bottom, we show unstable disks with active VSI, MRI and GI. From left to right, different disk inclinations are shown. The beam size is displayed at the bottom left corner of each panel. A $5\sigma$ clipping was applied to all moment maps. The color bar limits have been adjusted for each instability.}
\label{fig:nonkeplerianconvolved}
\end{figure*}

\subsection{Kinematic signatures in the raw images}\label{rawkinematicsignatures}

In the left $2.5$ panels of Figure \ref{fig:chanelmaps}, we show a set of raw channel maps computed from our radiative transfer models for a disk inclination of $30^{\circ}$. From top to bottom, we show the raw channels of disks unstable to the VSI, the MRI, and the GI. In all channels, perturbations are visible as spurs or wiggles at the edges of the emission areas, driven by the large-scale velocity perturbations. While in our models there are apparent differences between the signatures of the MRI and the GI in the channel maps, these signatures alone may not be sufficient to robustly distinguish their origin. In particular, the differences in the channel maps may become less distinguishable in the case of disks with stronger MRI or weaker GI than in our simulations.  
For mid-inclination disks, although both instabilities, the MRI and the GI,  show clear perturbations in all channel maps, the vigorous radial gas motions of the GI create substantial perturbations in the channels mapping near the systemic velocity, for which the GI produces its most studied signature, the `GI-wiggle' along the disk semi-minor axis \citep{Hall2020,Longarini2021b, Terry2022, Speedie2024}, whose quantitative analysis may hint the GI origin of the signatures.
However, in disks with low inclination, for which the radial velocity does not significantly contribute to the line-of-sight velocity, both the MRI and GI may produce similar features in the velocity channel maps. In any case, robust identification of the origin instability from the channel maps would require a framework to quantitatively study their properties across the full $\rm ^{12}CO$ line. Currently, the identification of the origin of the observed perturbations in the channel maps requires dedicated numerical simulations and radiative transfer predictions, such as it is the case of massive planets driving a localized signature \citep[see, e.g,][]{Pinte2023}, or GI-signatures \citep{Paneque2021, Speedie2024}.

In Figure \ref{fig:nonkeplerian}, we show the perturbations in the line centroid maps, predicted for our raw  synthetic images assuming a perfect background rotation subtraction. To obtain these ideal residuals, we generated a radiative transfer model in which we subtract the disk rotation in the input azimuthal velocity. In order to achieve a superb background rotation subtraction, we fit a power-law profile to the azimuthal average of the azimuthal velocity at each radius and height ($r$ and $\theta$; see also Section \ref{simulations}). The fitted 2D background rotation of the disk is then subtracted from the original azimuthal velocity field.
The line centroid velocity map of such a model directly gives the disk velocity residuals relative to its bulk rotation, showing the expected signatures from the perturbations driven by the disk instabilities. From top to bottom in Figure \ref{fig:nonkeplerian}, we show the residuals from our VSI-unstable, MRI-unstable, and GI-unstable disk models. From left to right, we show the effect of increasing disk inclination from $5^{\circ}$ to $60^{\circ}$, approximately covering the range of inclinations of the exoALMA sample of protoplanetary disks \citep[see Table 1 in][]{Teague_exoALMA}.

For low to moderate disk inclinations ($5^{\circ}\leq i \leq 30^{\circ}$), we recover the quasi axisymmetric rings and arcs in the residual maps from the VSI, predicted from the presence of strong meridional motions \citep{Barraza2021}, while for the MRI and the GI spiral-like structures dominate the line centroid residuals (see also \citealt{Hall2020}). For disk inclinations larger than $30^{\circ}$, the characteristic morphology of the residuals from the VSI is lost, and spiral-like arc structures are seen. The lower degree of azimuthal symmetry of VSI-induced non-Keplerian residuals for higher disk inclinations is a result from the larger contributions of the radial and azimuthal velocity components into the projected line of sight velocity. In particular, coherent radial velocity flows will have their strongest contribution to the line of sight along the semi-minor axis. In contrast, the contribution will be most substantial along the semi-major axis for the azimuthal velocity component \citep[see][]{Teague2019}. The distinct radial and azimuthal velocity structure induced by the VSI (see Figure \ref{fig:simulationssurface}) combined with the modulation of their contribution into the line-of-sight velocity varying disk PA, and partially mixed with the azimuthally-extended meridional flows, give rise to the observed pattern in the non-Keplerian residuals of the mid-inclination VSI-unstable disk. Therefore, the observed pattern in the residuals of the inclined VSI-unstable disk is mostly produced by the varying modulations of the radial and velocity components into the line of sight for different disk PA.

In the case of the MRI, higher disk inclinations strengthen the contribution from the sub- and super-Keplerian rings induced at the MRI dead-zone outer edge (see Section \ref{simulationsresults}).
In addition, non-axisymmetric structures seen as localized perturbations are apparent in the residuals of the mid-inclination disks ($30^{\circ}\leq i \leq 60^{\circ}$). Due to the projection effects, the azimuthal velocity deviations induced by non-Keplerian rings are more prominent along the disk's major axis, with opposite signs at the east and west sides of the disk. Similarly, the spiral-like radial velocity flows produced by the MRI are better traced for mid-inclination disk, inducing filamentary substructures. Still, the radial and azimuthal velocity perturbations are mixed when projected into the line of sight, in which their contributions vary depending on the disk PA, resulting in intricate non-Keplerian velocity residuals.

For the GI, the spiral velocity perturbations have similar morphology for all disk inclinations, due to the predominance of spiral structures in all of the three gas velocity components. Interestingly, fewer spiral arms are apparent for larger disk inclinations, resulting from the minor contribution to the line-of-sight velocity of the spatially narrower meridional spiral flows. Additionally, the residuals' strength is fairly consistent for all disk inclinations along the northern part of the disk minor axis, which can be attributed to the strong radial velocity of the spirals driven by the GI \citep{Hall2020, Longarini2021b, Speedie2024}. For the disk with an inclination of $60^{\circ}$, we attribute to projection effects along the line of sight the observed stronger residuals along the northern disk minor axis relative to the southern side of the disk, mainly due the inclined viewing angle combined with the elevation of the emission surface being traced.

\subsection{Signatures in the synthetic observations}\label{convolvedmodels}

In the right $2.5$ panels of Figure \ref{fig:chanelmaps}, we show the effect of exoALMA observation resolution in the $\rm ^{12}CO(3-2)$ channel maps (see details in Section \ref{methodscolvolved}), computed from our radiative transfer models for a disk inclination of $30^{\circ}$. Due to the effect of the synthetic observation resolution, the spurs and wiggles induced by the instabilities are blurred. In particular, the perturbations driven by the VSI and the MRI are challenging to identify, which may limit its observability in a direct inspection of the channel maps \citep[e.g.,][]{Pinte_exoALMA}. 

In Figure \ref{fig:nonkeplerianconvolved}, we show the kinematic signatures retrieved from synthetic observations with a spatial and spectral resolution comparable to that of the exoALMA survey. Following an observers' approach, the gas velocity perturbations are extracted by subtracting the line-of-sight velocity map of a smooth modified Keplerian disk model from the line-of-sight velocity map obtained from the unstable disks' synthetic observations (see details in Section \ref{kinematictools}). The resulting residuals unravel the kinematic signatures observable in exoALMA quality data. Relative to the unconvolved ideal case, presented in Figure \ref{fig:nonkeplerian}, the convolution by the Gaussian beam blurs the finer substructures driven by the instabilities, due to their spatial scales being smaller than the angular resolution (35 mas). Still, large-scale ring-like, arc-like, and spiral-like substructures are detectable and well-resolved with the exoALMA spatial resolution. Thus, the superb spectral resolution achieved by exoALMA, equal to a velocity resolution of $\sim 100\rm \, m\,s^{-1}$ in the $\rm ^{12}CO(3-2)$ fiducial images, is sufficient to clearly resolve in velocity the perturbations induced by the instabilities, capturing even subtle gas motions of a few tens of m s$^{-1}$.

In the predictions of the VSI, subtle perturbations with a quasi axisymmetric ring or an azimuthally extended arc-like morphology are recovered for low to intermediate inclinations. For higher disk inclinations ($i\geq 45^{\circ}$), the residuals have a more significant degree of asymmetry, resulting in arc-like features with smaller azimuthal extent. Additionally, the near side of the disk with an inclination of $i=60^{\circ}$ shows substantial systematic errors arising from the \textsc{eddy} model not accounting for radiative transfer effects in the line emission projected into the line of sight (also apparent in the MRI disk model with $i=60^{\circ}$). In particular, differences in the height of the probed $\rm ^{12}CO$ layer between the north and south regions of the disk are not accounted for in the \textsc{eddy} model.

For the MRI case, spiral-like structures are apparent for low disk inclinations ($i\leq 15^{\circ}$), while only filaments tracing spiral segments are recovered in our mid-inclination models ($30^{\circ}\leq i\leq 60^{\circ}$). We observe that, in some cases, spiral-like substructures that are separated features in the raw predictions (Figure \ref{fig:nonkeplerian}) are merged by the limited spatial resolution, appearing as one coherent large-scale structure, which can allow its observability within the capabilities of ALMA. For the disks with inclinations of $i=45^{\circ}$ and $i=60^{\circ}$, the dominant residuals are the azimuthal velocity deviations induced at the edge of the dead-zone, prominent along the semi-major axis of the disk. However, we stress that spatially resolved substructures such as the ones in our predictions are only resolvable in the exoALMA disks' outermost regions. Therefore, the MRI dead-zone outer edge should be located farther from the star in the exoALMA disks than in our model to produce resolvable features. An MRI dead-zone outer edge at a larger radius is a possibility, as its location can substantially vary depending on the systems' properties \citep{Delage2022}.
Last, localized features become more apparent with increasing disk inclination, with a prominent `Doppler-flip' resolved for the disk with $i=60^{\circ}$, which we discuss in Section \ref{localizedMRI}. 

The velocity perturbations recovered from our GI predictions are more prominent than those of the VSI or the MRI and, therefore, have a higher chance of detection within the exoALMA sample. The difference comes from the GI-driven spiral perturbations having a larger spatial scale and substantially higher velocities in our models. Our predictions indicate that large-scale spiral structures triggered by GI should be apparent if present in the sample, reaching magnitudes of $\sim 100$ m s$^{-1}$. These results align with estimations by analytical models of GI-induced spirals \citep{Longarini2021b} and are consistent with findings of GI signatures in AB Aur \citep{Speedie2024}.
For high disk inclinations ($i\geq 45^{\circ}$), our fitted disk rotation model to the GI-unstable disk requires the inclusion of a component that accounts for the super-Keplerian nature of massive self-gravitating disks (see Section \ref{kinematictools}). If the modification to the purely Keplerian rotation model is not included, a global dipole pattern dominates the line centroid map residuals. The need for this correction is consistent with other studies showing that, for fairly massive disks (with masses above $5\%$ the stellar mass), the effect of disk self-gravity in the disk rotation is detectable with high-resolution ALMA observations of molecular lines, allowing its kinematic mass measurement \citep{Veronesi2021, Veronesi2024,Andrews2024, Longarini_exoALMA}.

We also explored the non-axisymmetric signatures in the line peak intensity (or brightness temperature) and width (see method details in Section \ref{kinematictools}). The residuals in temperature and line width are shown in Figures \ref{fig:tempresiduals} and \ref{fig:widthresiduals} of the Appendix, respectively. The GI is the only instability that drives significant non-axisymmetric substructures in the temperature residuals. These deviations are likely due to changes in the elevation of the $\rm ^{12}CO$ emitting layer resulting from our simplified photo-dissociation prescription. Photo-dissociation removes CO deeper into the disk in the inter-arm region of the spirals. Because of the assumption of an smooth axisymmetric temperature structure in our model, the variations in CO emission height are reflected in brightness temperature deviations, traced by the peak intensity of the optically thick $\rm ^{12}CO$ line. Still, similar features may be observable in GI-unstable protoplanetary disks, and possibly amplified due to GI spiral arm heating from shocks and compression \citep[e.g.,][]{Cossins2009}, ignored in our radiative transfer model. 
In the case of the MRI, previous numerical simulations have predicted that it can drive non-axisymmetric tempreature deviations \citep[e.g.,][]{Ross2018}, ignored in our predictions due to the locally isothermal equation of state used in our MHD simulation, and the assumption of an axisymmetric disk temperature in the post-processing. However, MRI-driven temperature deviations found in previous works usually remain small \citep[see e.g.,][]{Flock2013}, and are mainly due to local current sheets \citep[see e.g., Fig. 9 in][]{Flock2013}. Moreover, temperature deviations driven by the MRI are subject of uncertainties related to radiative cooling times and efficiency. Models considering a temperature structure from a radiative magneto-hydrodynamic simulation of an MRI-active disk will be able to properly assess the detectability of MRI-induced temperature perturbations in ALMA CO rotational line emission observations.

Coherent non-axisymmetric line width residuals with spiral-like morphology are also prominent in the GI case, possibly originating from a combined effect of temperature broadening, turbulent broadening, and opacity broadening at overdensities \citep{Hacar2016}. However, for the line width map residuals, only the disk with an inclination of $5^{\circ}$ is not substantially affected by artifacts resulting from our procedure. The subtraction of an azimuthal average does not consider projection effects in the line width map, which requires a comprehensive disk model \citep[e.g., as provided by the Discminer modeling framework,][]{Izquierdo2021a, Izquierdo_exoALMA}. Our predictions of observable GI-induced spiral signatures in the line width map are in agreement with ALMA observations and supporting models exhibiting evidence of GI acting in the AB Aurigae planet-forming disk \citep{Speedie2024}.

While the observable non-axisymmetric residuals induced by the MRI or the large-scale spirals driven by the GI could be confused by planetary signatures, the different scenarios can be disentangled by using information of the peak and width of the emission line, in combination with the use of multiple molecular tracers, which we expand in Section \ref{localizedMRI}. Overall, our results for the VSI and the GI cases are consistent with previous findings presented in \citet{Barraza2021} and \citet{Hall2020}, while we show for the first time that MRI-driven large-scale velocity perturbations can be detectable with ALMA CO kinematics.


\section{Discussion}\label{discussion}

\subsection{Turbulence signatures in exoALMA disks}

Observations of the kinematic structure of protoplanetary disks from the exoALMA program have revealed that signatures of disk perturbations are common \citep{Teague_exoALMA}. These perturbations are present as `kinks' in channel maps and coherent substructures in the line centroid, line peak, and line width maps \citep[e.g.,][]{Pinte2018, Teague2022}. In particular, signatures in channel maps and residuals in the line centroid maps of $\rm ^{12}CO(3-2)$ line observations unravel the disk gas velocity perturbations in detail, hinting at the physical processes occurring in the disk.

The nearest protoplanetary disk in the sample, V4046 Sgr, shows no evidence of large-scale signatures in its outer regions \citep{Teague_exoALMA}, displaying smooth channel maps \citep{Pinte_exoALMA}. Similarly, smooth outer regions are present in the PDS 66 disks \citep{Teague_exoALMA, Pinte_exoALMA}. These results indicate that V4046 Sgr and PDS 66 have nearly laminar outer regions. Therefore, these regions have the specific conditions not to develop or sustain any disk instability: they have low surface densities such that they are stable to the GI, they have a low ionization fraction such that the gas is not susceptible to magnetic effects, that is, stable to the MRI, and have cooling time-scales sufficiently large to damp VSI-motions due to the disk vertical buoyancy \citep[time-scales comparable to or larger than the local dynamical time-scale $\Omega_{\rm Kep}^{-1}$,][]{Nelson2013, Lin2015}. These results are consistent with previous turbulence level upper limits, constraining weak turbulence in the outer regions of V4046 Sgr \citep{Flaherty2020}.

The quasi axisymmetric ring and arc-like non-Keplerian motions predicted from large-scale VSI-driven turbulence are present in two disks from the exoALMA sample \citep[see, e.g.,][]{Izquierdo_exoALMA}. In particular, LkCa 15 shows arcs of non-Keplerian motions in the line centroid map residuals \citep{Izquierdo_exoALMA}, and azimuthal velocity deviations in its rotation curve in the rotation curves \citep{Stadler_exoALMA, Izquierdo_exoALMA}. However, its inclination of $\sim 50^{\circ}$ makes a challenging case to robustly identify VSI motions due to the minor contribution of the meridional velocity into the line-of-sight velocity (see Section \ref{rtresults}). Similarly, the outermost regions of RXJ 1604 show one ring-like structure in its projected velocity residuals \citep{Stadler2023}. However, its apparent connection to a tightly wound spiral might suggest a planetary origin \citep{Stadler2023}. In addition, predictions of VSI motions result in multiple concentric rings or arcs of Doppler-shifted velocities, not a single ring-like substructure. The overall lack of VSI motions in the exoALMA disks could indicate that most disks have relatively long cooling time-scales in their outer regions, which halts the development of the VSI. Alternatively, it could suggest that the gas dynamics of the disk in the sample is dominated by other disk instabilities (e.g., the MRI or the GI) or planet--disk interactions.

Spiral structures are present in several exoALMA disks, including CQ Tau, HD 135344B, and MWC 758 \citep{Wolfer2021, Wolfer_exoALMA, Izquierdo_exoALMA, Hilder_exoALMA}. GI- or MRI-driven motions could be responsible for some of the observed spiral-like velocity residuals. In particular, GI-induced substructures could also explain some of the observed spiral features in temperature and line width \citep{Izquierdo_exoALMA, Wolfer_exoALMA}. However, the strong non-axisymmetric features or low disk inclinations impede the accurate measurement of the kinematic disk mass of the targets with the most prominent large-scale spirals \citep[including MWC 758, CQ Tau, HD 135344B, HD 143006 and J1604,][]{Longarini_exoALMA}. Therefore, their stability against GI is still an open question. The subset of disks with kinematic mass measurements is GI-stable \citep[][also validated by an alternative method in \citealt{Trapman_exoALMA}]{Longarini_exoALMA}. Therefore, only the MRI could still be active within that subset of disks, driving some of the filamentary spiral-like structures observed.

The only disk in the exoALMA sample with previous measurements of non-zero turbulence, DM Tau, shows evidence of large-scale flows in the channel maps \citep[e.g.,][Hardiman et al. in prep.]{Pinte_exoALMA}. Its disk mass measured via kinematics indicates that the DM Tau is GI-stable \citep{Longarini_exoALMA}, indicating the possibility of active MRI, previously proposed as the source of the observed non-thermal broadening \citep{Flaherty2020}. Nevertheless, efforts in constraining the level of turbulence in DM Tau via line broadening from the exoALMA data have shown that these measurements are still degenerated even for high quality data (Hardiman et al. in prep.). Particularly, the level of micro-scale turbulence needed to fit the data is highly sensitive to the model temperature structure and the prescription used for the parameter controlling the turbulent broadening (Hardiman et al. in prep.; see also \citealt{Flaherty2015}). Therefore, new methods to measure turbulence that lift these degeneracies and can account for resolved substructures are needed to robustly measure the turbulent broadening in DM Tau (Hardiman et al. in prep.). Further, analysis of the residual maps and models of higher complexity are required to obtain conclusions regarding the nature of turbulence in DM Tau, possibly confirming the MRI-origin of its large-scale perturbations.

The disks around HD 34282 \citep{Wolfer_exoALMA} and J1615 \citep{Galloway_exoALMA} show an intricate velocity structure. Similar to the case of LkCa 15, their inclinations of $\sim 58^{\circ}$ and $\sim 47^{\circ}$, respectively, make it difficult to interpret their line centroid residuals due to the mixed contributions of the different disk velocity components into the line-of-sight velocity. A forthcoming paper will present a quantitative analysis of the 2D residual maps and their comparison to predictions from large-scale turbulence for the complete exoALMA sample.

Finally, several exoALMA targets show evidence of dynamical interactions with massive planets or companions \citep{Pinte_exoALMA}. The perturbations driven by massive planets can drive strong signatures \citep[e.g.,][]{Wolfer2021, Izquierdo_exoALMA}, complicating the identification of substructures from other physical mechanisms. Therefore, planet--disk interactions can make it more difficult to characterize signatures of turbulent motions \citep[e.g.,][]{Barraza2024}. Follow-up higher spatial resolution observations exploiting the spectral resolution available with ALMA, combined with simultaneous predictions of velocity, temperature, and line width of disk perturbations, will lead to the robust characterization of large-scale turbulence signatures in exoALMA disks.

\begin{figure*}[htp]
\centering
\includegraphics[angle=0,width=\linewidth]{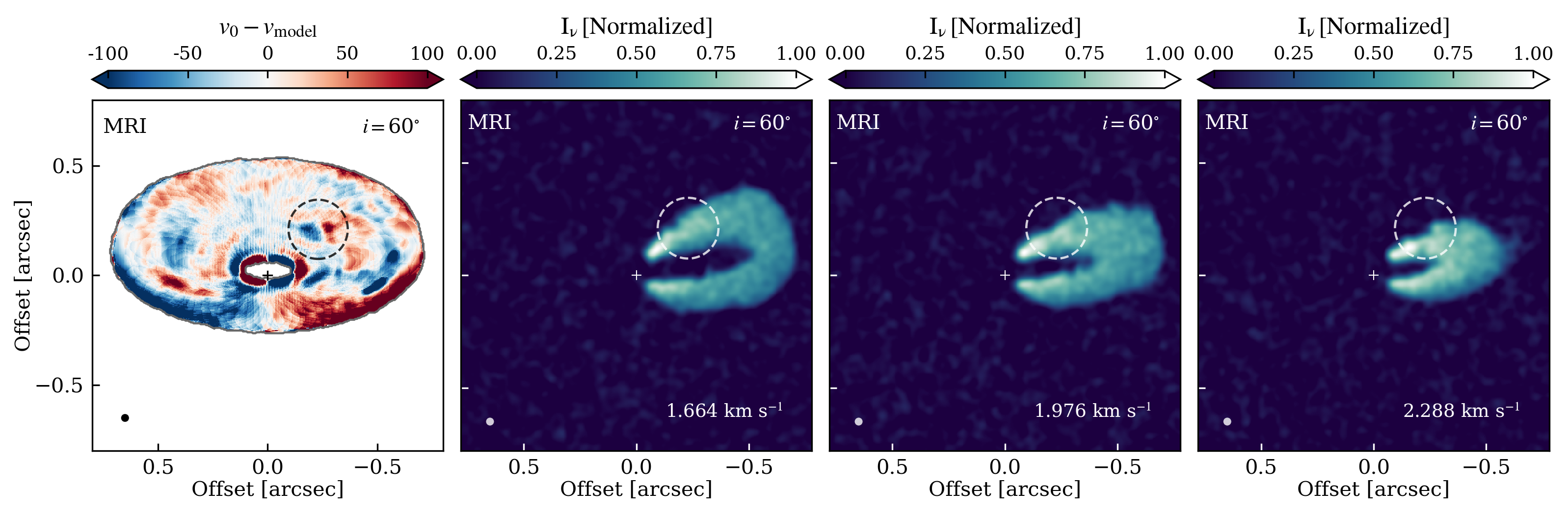}
\caption{$\rm ^{12}CO(3-2)$ line centroid map residuals and selected channel maps from synthetic predictions of an MRI-unstable disk for a disk with an inclination of $60^{\circ}$, highlighting a localized feature. The images include the effect of the spatial resolution ($35$ mas FWHM Gaussian beam, shown in the bottom left corner), spectral resolution ($104$ m s$^{-1}$ channel spacing), and sensitivity equivalent to that of exoALMA.}
\label{fig:localizedfeatureMRI}
\end{figure*}

\subsection{Localized features from MRI-driven perturbations}\label{localizedMRI}

Localized features (in radius and azimuth) are often interpreted as the presence of embedded massive planets due to their strong influence around their location. An embedded massive planet can manifest in the observables as a major `kink' spanning a set of channel maps \citep[e.g.,][]{Pinte2019, Pinte_exoALMA, Bae_exoALMA}, a `Doppler-flip' (dipole pattern of blue- and red-shifted velocities) in the line centroid residuals from Keplerian rotation \citep[e.g.,][]{Perez2018,Casassus2022, Teague2022, Bae_exoALMA}, and a local excess in the line width around its location \citep{Izquierdo_exoALMA}.
Interestingly, our $\rm ^{12}CO(3-2)$ predictions of kinematic signatures from MRI-driven motions show a localized feature (highlighted in Figure \ref{fig:localizedfeatureMRI}). In particular, the line centroid map residuals for the model with an inclination of $i=60^{\circ}$ shows a localized signature at roughly $\varphi = 45^{\circ}$ west of north, seen in the rightmost panel of the middle row of Figures \ref{fig:nonkeplerian} and \ref{fig:nonkeplerianconvolved}. Moreover, the localized velocity perturbation creates apparent wiggles in a limited range of channel maps.
While we do not thoroughly compare such features with planet-driven velocity deviations, our predictions indicate that MRI-driven turbulence can create localized features that could be misinterpreted as embedded planets.
However, we observe that MRI-driven perturbations substantially change in location, morphology, and strength with disk height. Therefore, the MRI origin of localized features can be ruled out if the detected signatures in channel maps and line centroid residuals are consistent among different tracers, as they probe distinct disk layers. 
In addition, our synthetic predictions do not show an excess of the $\rm ^{12}CO(3-2)$ line width co-located with the MRI-induced localized feature (see Figure \ref{fig:widthresiduals} in the Appendix), different from the predictions of signatures induced by embedded massive planets. Yet, the line width map residuals of the mid-inclination disks are affected by artifacts, limiting the observability of the velocity-induced broadening from the localized feature.
Our results highlight the importance of observing and analyzing localized perturbations in multiple molecular tracers, as performed by the exoALMA large program. Furthermore, our predictions stress the crucial role of alternative observables to the line centroid to robustly identify planet-driven signatures, such as the line peak and line width extracted with line profile tomography \citep{Izquierdo_exoALMA}. Still, a systematic study directly comparing the observability of localized kinematic signatures from MRI and embedded planets in molecular line observations is needed to confirm the hinted features that could be used to distinguish between scenarios. 

\subsection{Limitations of our numerical methods}\label{caveats}

Global 3D (magneto-)hydrodynamical simulations that include additional physics could provide new insights into the expected kinematic structure of magnetized protoplanetary disks. In particular, our post-processed non-ideal MHD simulations ignore the effect of the Hall effect and ambipolar diffusion. Ambipolar diffusion can suppress the MRI activity in the outer disk \citep[see][and references therein]{Lesur2023}, while both ambipolar diffusion and the Hall effect can induce additional large-scale substructures \citep[e.g.,][]{Bethune2016, Cui2021}. These non-ideal MHD effects can also modify the extent of the MRI dead zone, and influence the gas dynamics interior to the MRI dead-zone and near its edges \citep{Kunz2013, Lesur2014, Delage2022, Iwasaki2024}. Moreover, the strength of the ambipolar diffusion can define if the MRI or the VSI dominates in outer regions of fast-cooling GI-stable disks \citep{Cui2022}. Also, we neglect the magnetic effects on the GI gas dynamics in our SGHD simulation. In magnetized disks, the GI turbulent motions can drive a dynamo effect amplifying the magnetic field, resulting in a saturated state with less defined spiral arms \citep{Deng2020, Bethune2022}, possibly more difficult to identify and characterize in CO kinematic observations.

The simulations presented in this paper are fine-tuned for the detailed study of disk instabilities; therefore, they are not tailored toward specific protoplanetary disks. Additionally, our models are set to explore a limited radial region of the disk (from $20$ au to $100$ au), while the observed disks can extend up to several hundreds of au in size \citep[e.g.,][]{Galloway_exoALMA, Longarini_exoALMA}.
Further studies on identifying kinematic signatures from turbulence in the exoALMA sample will require source-specific numerical simulations and modeling, including information on each disk's structure. For example, we could use the disk temperature structure, stellar dynamical mass, disk size, disk mass, and other disk properties constrained from the exoALMA data \citep[see, e.g.,][]{Galloway_exoALMA,Longarini_exoALMA}. In particular, the local pressure scale height, set by the disk temperature and increasing with disk radius, will define the radial size of substructures \citep{Bae2023}; therefore, it will set their observability within the spatial resolution of exoALMA. These new constraints on disk properties from the exoALMA observations add valuable information to guide a parameter space exploration in 3D hydrodynamical simulations, which will substantially help conclude if the disk has the conditions to develop and sustain the steady state of the disk instabilities. By confirming that disk instabilities can drive the observed substructures unraveled via line profile tomography, we will ultimately have a robust detection of a turbulence mechanism active in planet-forming disks \citep[e.g.,][]{Speedie2024}.

\subsection{Effect of systematic errors in the characterization of turbulence-driven substructures}

We have presented predictions of coherent kinematic substructures driven by large-scale motions from disk instabilities. The observational signatures are extracted by subtracting a smooth background model for the disk rotation in the observed line centroid maps. However, such an approach is sensitive to various systematic errors, and careful extraction of the residuals is a must. First, small errors in the fitting parameters of a (sub- or super-)Keplerian model to the disk line centroid map obtained from rotational line observations can lead to global modulations \citep{Yen2020, Norfolk2022}. In particular, patterns with characteristic dipole or quadrupole patterns will appear in the line centroid residual maps, prominent at the innermost and outermost regions of the disk. Such an artifact in the residual maps can introduce limitations on extracting coherent substructures, or errors in any analysis of the residual maps properties. Ideally, methods to determine disk regions affected by systematic errors have to be applied during analysis to avoid considering them when interpreting the observations. In this direction, the data analysis tools developed within the exoALMA collaboration provide a significant step in advancing towards robust frameworks to extract observables and their residuals accurately \citep[see, e.g.,][]{Izquierdo_exoALMA,Hilder_exoALMA}. In addition, a comprehensive exploration of the kinematic analysis tools has been performed as part of exoALMA for post-processed 3D numerical simulations of planet--disk interactions, whose benchmarking will be fundamental for future studies of kinematic signatures \citep{Bae_exoALMA}.

\subsection{Influence of embedded planets}

The simulations presented in this work do not include the presence of embedded planets or companions in the disk. The presence of massive planets strongly modifies the disk's dynamical structure, inducing large-scale spiral arms, rings of sub- and super-Keplerian velocities, meridional flows, and vortices \citep[for a review see][and references therein]{Paardekooper2023}. Additional kinematic substructures driven by planets can alter the moment maps residuals \citep{Izquierdo_exoALMA}, affecting the extraction of information from active turbulence. In particular, it has been demonstrated that massive planets can suppress observational signatures of the GI \citep{Rowther2020} and the VSI \citep{Stoll2017, Ziampras2023, Hammer2023, Barraza2024}. 
Similarly, ideal magneto-hydrodynamic simulations have shown that a massive planet embedded in the disk can reduce the magnetic stresses, possibly weakening the MRI-driven turbulence along the gap region \citep{Nelson2003, Uribe2011, Cilibrasi2023}. Moreover, the inclusion of non-ideal MHD effects and magnetic winds in magnetized planet--disk interaction simulations results in a complex disk dynamical structure \citep{Aoyama2023, Wafflard2023, Hu2025}, potentially hiding kinematic signatures of the MRI.
Still, to properly address the influence of planets in the disk structure, a precise knowledge of the disk's physical properties and the location and masses of embedded planets is needed  \citep[see, e.g.,][]{Pinte_exoALMA}. To this end, the constraints obtained from exoALMA will help to conduct dedicated simulations, exploring the kinematic signatures resulting from the interplay of turbulence-driving mechanisms and embedded planets. 


\section{Summary and Conclusions}\label{conclusions}

We conducted a study of the large-scale turbulence signatures from numerical simulations of unstable protoplanetary disks. In particular, we explored how coherent substructures are manifested in the residuals from a background disk model in exoALMA quality data of $\rm ^{12}CO(3-2)$ line emission.
Post-processing 3D global (magneto-)hydrodynamic simulations with radiative transfer calculations, we examined three disk instabilities candidates to operate in the outer regions of planet-forming disks: the vertical shear instability (VSI), the magneto-rotational instability (MRI), and the gravitational instability (GI). From building a gallery of expected outcomes, we put our results in the context of the exoALMA survey, discussing the qualitative resemblance of the perturbations extracted in exoALMA disks and our predictions of substructures driven by disk instabilities. We summarize our results in the following:

\begin{itemize}
    \item Our results show that the large-scale perturbations driven by the VSI, MRI, and GI are detectable by exoALMA quality $\rm ^{12}CO(3-2)$ observations. Among the instabilities, the VSI shows the weakest perturbations with dominant ring- and arc-like substructures in the velocity centroid map residuals from a (sub-)Keplerian disk, decreasing in their degree of axisymmetry with increasing disk inclination. 
    \item Velocity perturbations driven by the MRI manifest as spiral-like substructures. Additionally, for mid-inclination disks, non-Keplerian rings at the MRI dead-zone outer edge are apparent. Similarly, the GI drives large-scale coherent substructures with spiral morphology, present in the residuals of line centroid, line peak, and line width.
    \item We predict that the MRI can induce localized kinematic signatures in mid-inclination disks. These signatures may resemble the signposts induced by embedded massive planets. However, the MRI localized signature is observed only in a layer of the disk. Therefore, observations of multiple molecular tracers can help distinguish between scenarios.  
    \item By qualitatively comparing the overall morphology of perturbations from our predictions and the results with that from exoALMA observations, a robust case of VSI acting in the outermost regions of the disk is not yet observed.
    \item The non-detection of spurs in the channel maps of PDS 66 and V4046 Sgr implies that their outer regions are nearly laminar. Therefore, suggesting that they are weakly ionized (stable to the MRI), have low gas surface densities (stable to the GI), and have relatively long cooling time-scales (stable to the VSI). At the other end, dynamic exoALMA disks, such as MWC 758, HD 135344B, and CQ Tau, show apparent substructures with spiral morphologies, indicating that spiral-inducing mechanisms dominate (e.g., the MRI, the GI, or massive planets).
    \item Spirals driven by the MRI or the GI could explain some of the substructures observed in exoALMA targets. However, these disks show evidence of dynamical interactions with embedded massive planets, resulting in a complex kinematic structure. Further work addressing the presence of disk instabilities from a methodical study of the statistical properties of residual maps is required to disentangle between scenarios.
\end{itemize}

\section*{Acknowledgments}

This paper makes use of the following ALMA data: ADS/JAO.ALMA\#2021.1.01123.L. ALMA is a partnership of ESO (representing its member states), NSF (USA) and NINS (Japan), together with NRC (Canada), MOST and ASIAA (Taiwan), and KASI (Republic of Korea), in cooperation with the Republic of Chile. The Joint ALMA Observatory is operated by ESO, AUI/NRAO and NAOJ. The National Radio Astronomy Observatory and Green Bank Observatory are facilities of the U.S. National Science Foundation operated under cooperative agreement by Associated Universities, Inc. We thank the North American ALMA Science Center (NAASC) for their generous support including providing computing facilities and financial support for student attendance at workshops and publications.

The authors thank the anonymous referee, whose feedback significantly improved the manuscript.
The authors acknowledge MIT Office of Research Computing and Data, the MIT SuperCloud and Lincoln Laboratory Supercomputing Center for providing HPC resources that have contributed to the research results reported within this paper. MF has received funding from the European Research Council (ERC) under the European Unions Horizon 2020 research and innovation program (grant agreement No. 757957). 
JB acknowledges support from NASA XRP grant No. 80NSSC23K1312. 
MB, DF, JS, and AJW have received funding from the European Research Council (ERC) under the European Union’s Horizon 2020 research and innovation programme (PROTOPLANETS, grant agreement No. 101002188).
Computations by JS have been performed on the `Mesocentre SIGAMM' machine, hosted by Observatoire de la Cote d’Azur. 
PC acknowledges support by the Italian Ministero dell'Istruzione, Universit\`a e Ricerca through the grant Progetti Premiali 2012 – iALMA (CUP C52I13000140001) and by the ANID BASAL project FB210003.
SF is funded by the European Union (ERC, UNVEIL, 101076613), and acknowledges financial contribution from PRIN-MUR 2022YP5ACE. 
MF is supported by a Grant-in-Aid from the Japan Society for the Promotion of Science (KAKENHI: No. JP22H01274). 
CH acknowledges support from NSF AAG grant No. 2407679.
JDI acknowledges support from an STFC Ernest Rutherford Fellowship (ST/W004119/1) and a University Academic Fellowship from the University of Leeds. 
Support for AFI was provided by NASA through the NASA Hubble Fellowship grant No. HST-HF2-51532.001-A awarded by the Space Telescope Science Institute, which is operated by the Association of Universities for Research in Astronomy, Inc., for NASA, under contract NAS5-26555.
GL and GWF acknowledge support from the European Research Council (ERC) under the European Union Horizon 2020 research and innovation program (Grant agreement no. 815559 (MHDiscs)).
CL has received funding from the European Union's Horizon 2020 research and innovation program under the Marie Sklodowska-Curie grant agreement No. 823823 (DUSTBUSTERS) and by the UK Science and Technology research Council (STFC) via the consolidated grant ST/W000997/1. 
CP acknowledges Australian Research Council funding via FT170100040, DP18010423, DP220103767, and DP240103290. DP acknowledges Australian Research Council funding via DP18010423, DP220103767, and DP240103290. 
GR acknowledges funding from the Fondazione Cariplo, grant no. 2022-1217, and the European Research Council (ERC) under the European Union’s Horizon Europe Research \& Innovation Programme under grant agreement no. 101039651 (DiscEvol). 
H-WY acknowledges support from National Science and Technology Council (NSTC) in Taiwan through grant NSTC 113-2112-M-001-035- and from the Academia Sinica Career Development Award (AS-CDA-111-M03). GWF was granted access to the HPC resources of IDRIS under the allocation A0120402231 made by GENCI. 
AJW has received funding from the European Union’s Horizon 2020
research and innovation programme under the Marie Skłodowska-Curie grant agreement No 101104656.
TCY acknowledges support by Grant-in-Aid for JSPS Fellows JP23KJ1008.
Support for BZ was provided by The Brinson Foundation. 
Views and opinions expressed by ERC-funded scientists are however those of the author(s) only and do not necessarily reflect those of the European Union or the European Research Council. Neither the European Union nor the granting authority can be held responsible for them. 

\software{PLUTO \citep{Mignone2007},  
          RADMC-3D \citep{Dullemond2012}, 
          fargo2radmc3d \citep{Baruteau2019},
          CASA \citep{CASA},
          bettermoments \citep{bettermoments},
          emcee \citep{emcee},
          eddy \citep{eddy},
          GoFish \citep{GoFish},
          Numpy \citep{Numpy},
          Scipy \citep{Scipy},
          VTK \citep{vtkBook},
          Matplotlib \citep{Hunter2007},
          CMasher \citep{CMasher}
          }

\bibliography{references}{}
\bibliographystyle{aasjournal}

\appendix
\counterwithin{figure}{section}

\section{Eddy disk model and fitting procedure}\label{app:diskmodeleddy}

In our \textsc{eddy} disk models described in Section \ref{kinematictools}, we include a modified Keplerian rotation profile that includes a radially-varying deviation from purely Keplerian rotation, parametrized by a mass component \citep{Teague2022}:
\begin{equation}
    v_{\phi} = \sqrt{\frac{G(M_{\star}+ M_d(R)) R^2}{(R^2+z^2)^{3/2}}},
\end{equation}
\begin{equation}
    M_{d}(R) = M_{\rm disk} \times \frac{R^{2-\gamma}-R_{\rm in}^{2-\gamma}}{R_{\rm out}^{2-\gamma}-R_{\rm in}^{2-\gamma}},
\end{equation}
where $R$ is the cylindrical radius, $z$ is the disk surface height, and $M_{\star}$ is the dynamical mass of the central star. The parameters $R_{\rm in}$ and $R_{\rm out}$ describe the disk inner and outer edge, respectively. $M_{\rm disk}$ represents the total disk mass, and $\gamma$ is the exponent of the underlying power-law surface density profile, assumed to be of the form $\Sigma(r)=\Sigma_0\times R^{-\gamma}$.

The parametrization of the emission surface in our models is given by:
\begin{equation}
    z(R^{*}) = z_0 \times \left( \frac{R^{*}}{1\arcsec} \right)^{\psi}\times\exp{-\left(\frac{R^{*}}{R_{\rm taper}} \right)^{q_{\rm taper}}},
\end{equation}
where the cylindrical radius $R^{*}$ and scale radius of the exponential taper $R_{\rm taper}$ are implemented as angular distances defined in arcsecond. The disk aspect ratio of the emission surface is set by $z_0$, $\psi$ describes the surface flaring, and $R_{\rm taper}$ and $q_{taper}$ describe an exponential taper to the emission height at the disk outermost region. Finally, the fitting is done in the projected velocity into the line of sight $v_0$, described as:
\begin{equation}
    v_0 =v_{\phi}\cos{\phi} \sin{i} + v_{\rm LSR},
\end{equation}
where $\phi$ is the polar angle, $i$ the disk inclination, and $v_{\rm LSR}$ is the systemic velocity. 
The free parameters in our calculations are $M_{\star}$, the emission surface parameters $[z_0$; $\psi$; $R_{\rm taper}$; $q_{\rm taper}]$, the disk inclination $i$, the disk mass parameters $[M_{\rm disk}$; $\gamma]$, and the center of the image $[x_0, y_0]$. We fixed the disk position angle ${\rm PA}$ to the input model value, the systemic velocity $v_{\rm LSR}$ set to zero in our models, the distance to the source to $140$ pc, and the inner and outer disk edges considered in the disk mass component $r_{\rm in}$ and $r_{\rm out}$, set to $20$ and $100$ au, respectively.

To obtain our best fit disk models, we computed the posterior distribution of our free parameters with MCMC sampling implemented with the \texttt{emcee} package \citep{emcee} in \texttt{eddy}. We set 80 walkers initialized with the disk radiative transfer input model parameters for $M_{\star}$ and $M_{\rm disk}$, while $[\gamma=1; x_0=0, y_0=0]$ were used as initial values. For the emission surface parameters, an approximated location was obtained by roughly reproducing the layer where the optical depth reaches unity in our model (see Figure \ref{fig:numberdensmodels} of the Appendix). We set 2000 steps for the walkers to burn in, and 4000 additional to sample the posterior distribution.

\newpage
\section{Parameters of the simulations}\label{app:tables}

\begin{table}[h]
\centering
\begin{tabular}{ |c| c| c| c| }
Label & RHD [VSI] & MHD [MRI] & SGHD [GI]\\
\hline
$L_{\star}$ & $ 0.95\,L_{\odot}$ & $ 0.95\,L_{\odot}$  & $-$ \\
$R_{\star}$ & $ 2.0\,R_{\odot}$ & $2.0\,R_{\odot}$ & $-$ \\
$M_{\star}$ & $0.5\,M_{\odot}$ & $0.5\,M_{\odot}$ & $0.5\,M_{\odot}$   \\
$T_{\star}$ & $4000\,\textrm{K}$ & $4000\,\textrm{K}$ & $-$\\
Disk total gas mass & 0.05 $M_{\star}$ & 0.085 $M_{\star}$ & 0.2 $M_{\star}$\\
$\Sigma_0$ at $R=100$ au & 6.0 g cm$^{-2}$ & 5.94 g cm$^{-2}$ & $-$\\
Spectral type & K & K & $-$ \\
Dust-to-gas mass ratio  & $10^{-3}$ & $10^{-2}$ & $-$\\
Maximum dust size & 10 $\mu$m & 10 $\mu$m & $-$ \\
Minimum dust size & 0.01 $\mu$m & 0.01 $\mu$m & $-$ \\
Dust size slope & -3.5  & -3.5  & $-$ \\
$\Delta r$ & $20-100$ au & $20-100$ au & $5-160$ au\\
$\Delta \theta$ & $0.7$ rad & $0.72$ rad  & $0.7$ rad \\
$\Delta \phi$ & $2\pi$  rad & $2\pi$ rad & $2\pi$ rad\\
$N_r$ & $1024$ & $256$ & $518$ \\
$N_{\theta}$ & $512$ & $128$ & $96$\\
$N_{\phi}$ & $2044$ & $512$ & $512$\\
Reference $H/R$ & $0.1$ & $0.1$ & $0.028$\\
cells per $H$ & $70$ & $20$ & $^{\star}9$\\
$\langle \alpha \rangle$ & $1.5\times 10^{-4}$ & $\sim 3\times 10^{-3}$ & $9.7\times 10^{-2}$\\
\end{tabular} 
\caption{Summary of the simulations' parameters. See further details in \cite{Flock2015}, \cite{Flock2020} and \citealt{Bethune2021}, for the MHD, RHD, and SGHD simulations, respectively. Disk- and time-averaged values of the effective $\alpha$ viscosity \citep{Shakura1973} are reported. $^{\star}$Cells per pressure scale height near the disk midplane.} 
\label{tab:simulationsparameters}
\end{table}

\newpage
\section{Additional Figures}\label{app:additionalfigures}

\begin{figure}[htp]
\centering
\includegraphics[angle=0,width=\linewidth]{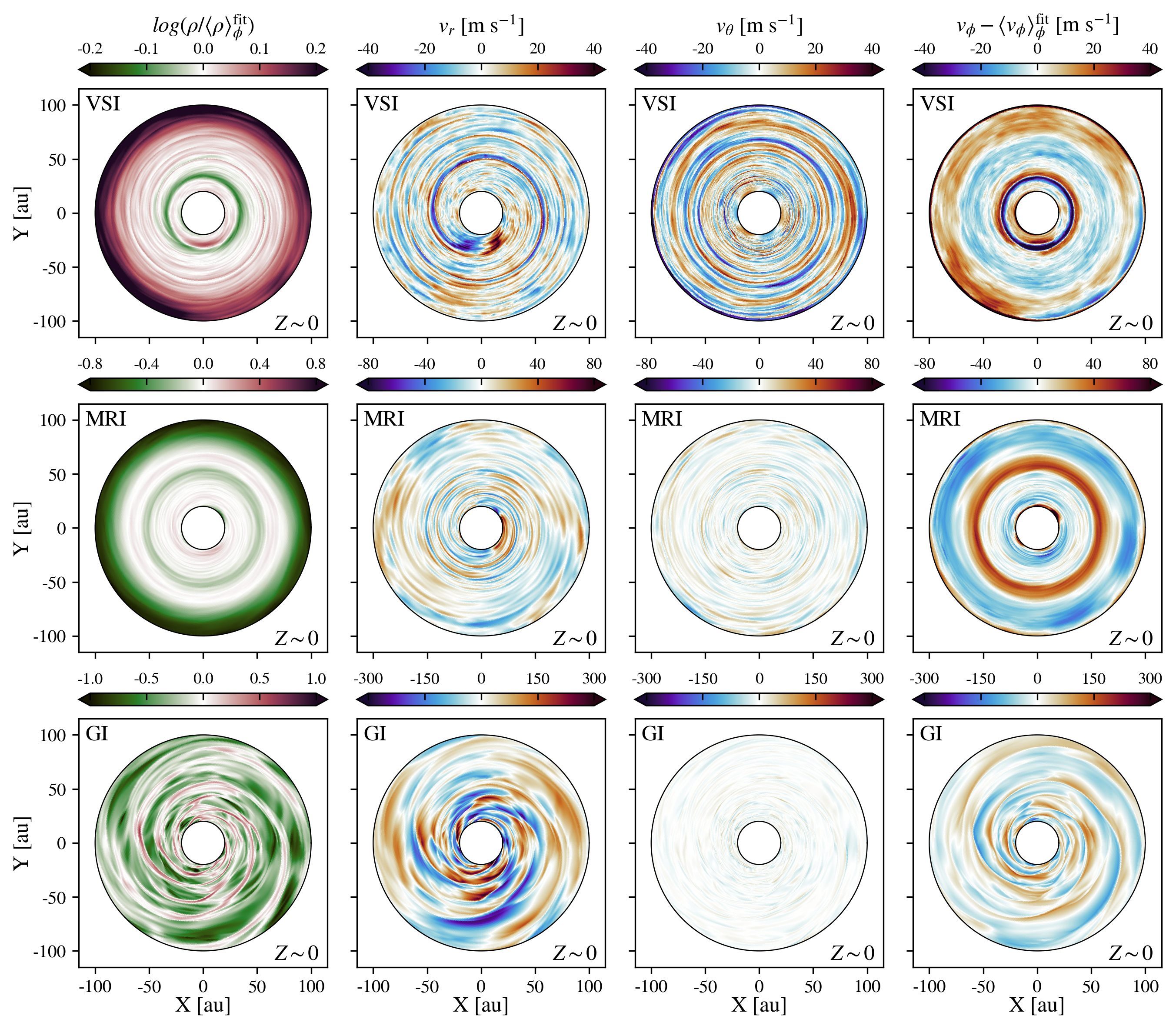}
\caption{Cartesian view of perturbations at the disk midplane for the HD (top), MHD (middle) and SGHD (bottom) simulations, with active VSI, MRI, and GI, respectively. From left to right, we show gas density perturbations, radial velocity, meridional velocity, and azimuthal velocity perturbations. The color bar limits are adjusted for each instability.}
\label{fig:simulationsmidplane}
\end{figure}

\begin{figure}[htp]
\centering
\includegraphics[angle=0,width=\linewidth]{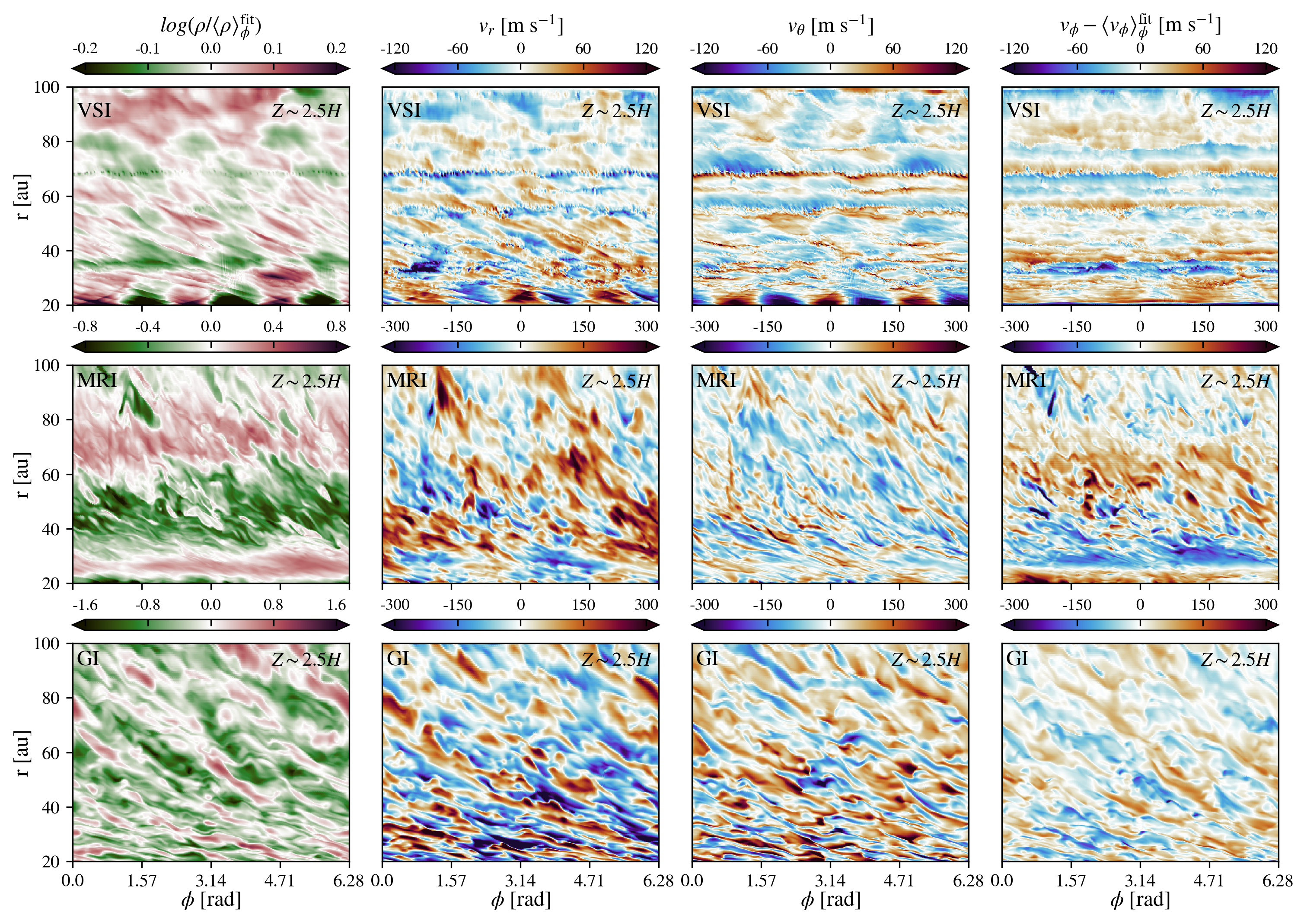}
\caption{Polar view of the disk perturbations at $Z\sim 2.5H$ for the HD (top), MHD (middle) and SGHD (bottom) simulations, with active VSI, MRI, and GI, respectively. From left to right, perturbations in density, radial velocity, meridional velocity, and azimuthal velocity. The color bar limits are adjusted for each instability.}
\label{fig:simulationspolar}
\end{figure}

\begin{figure}[htp]
\centering
\includegraphics[angle=0,width=\linewidth]{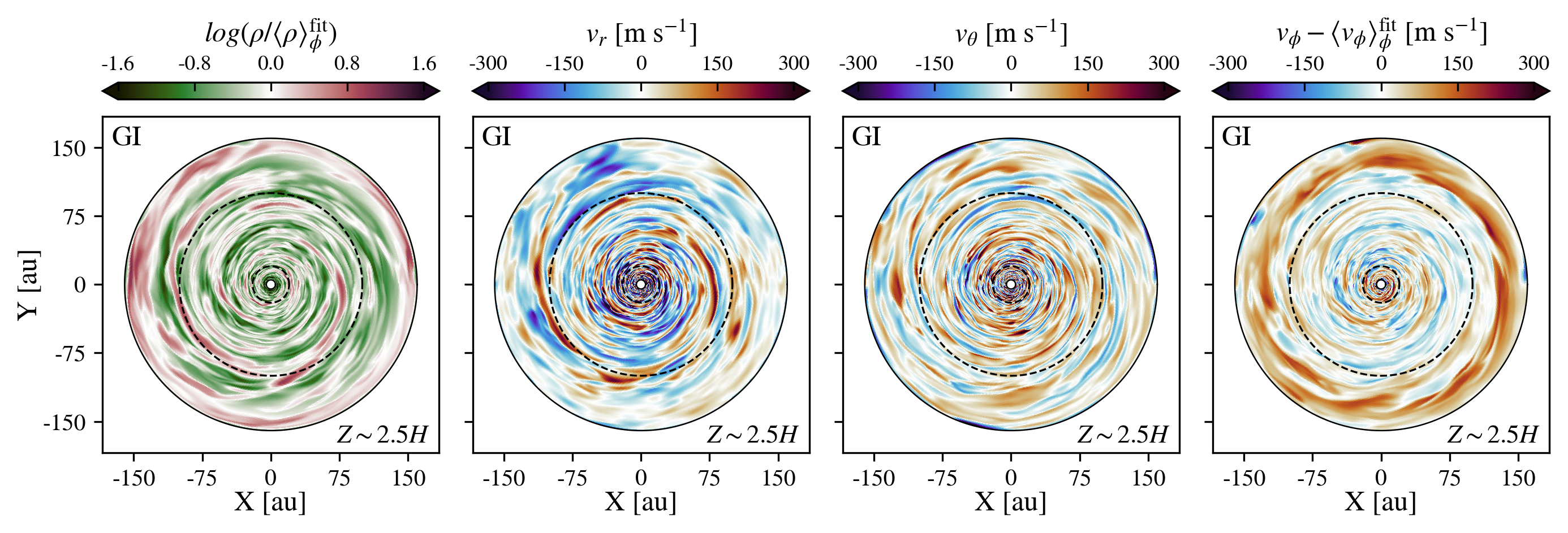}
\caption{Cartesian view of perturbations at $Z\sim 2.5H$ for the SGHD simulation with active GI, showing the simulation full domain. From left to right, gas density perturbations, radial velocity, meridional velocity, and azimuthal velocity perturbations. The dotted lines enclose the radial region of the simulation considered in our study.}
\label{fig:gisimulationfull}
\end{figure}

\begin{figure}[htp]
\centering
\includegraphics[angle=0,width=\linewidth]{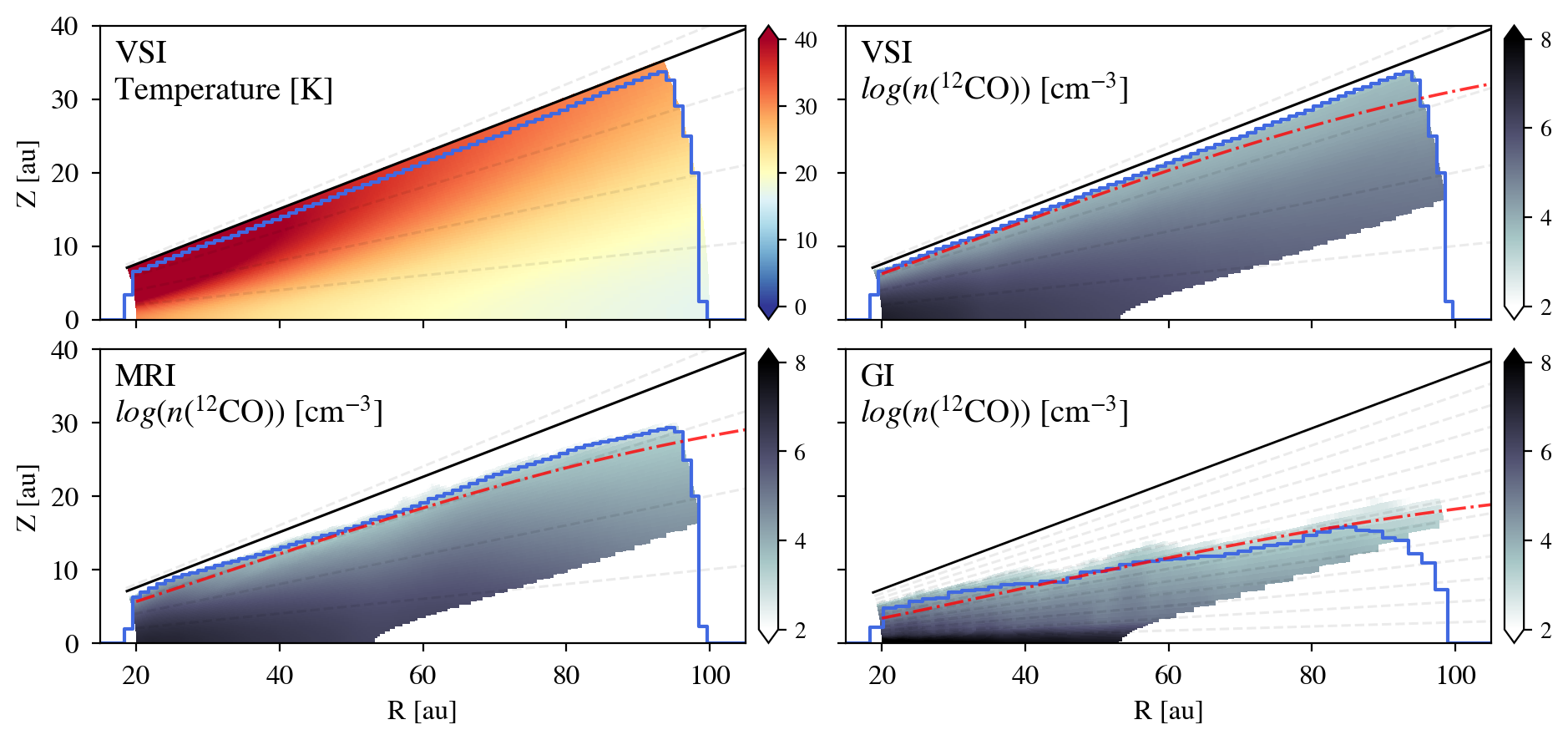}
\caption{Gas temperature and azimuthally averaged number density of $\rm ^{12}CO$ of our disk radiative transfer models. Overlaid in blue, a solid line traces the height at which the optical depth reaches unity assuming a face-on disk. The red dash-dotted line shows the initial guess for the $\rm ^{12}CO(3-2)$ emission surface in our \textsc{eddy} models. The solid black line indicate the simulations' domain in colatitude, while the grey dashed lines show spacings along $Z$ equal to the disks' pressure scale height.}
\label{fig:numberdensmodels}
\end{figure}

\begin{figure}[htp]
\centering
\includegraphics[angle=0,width=\linewidth]{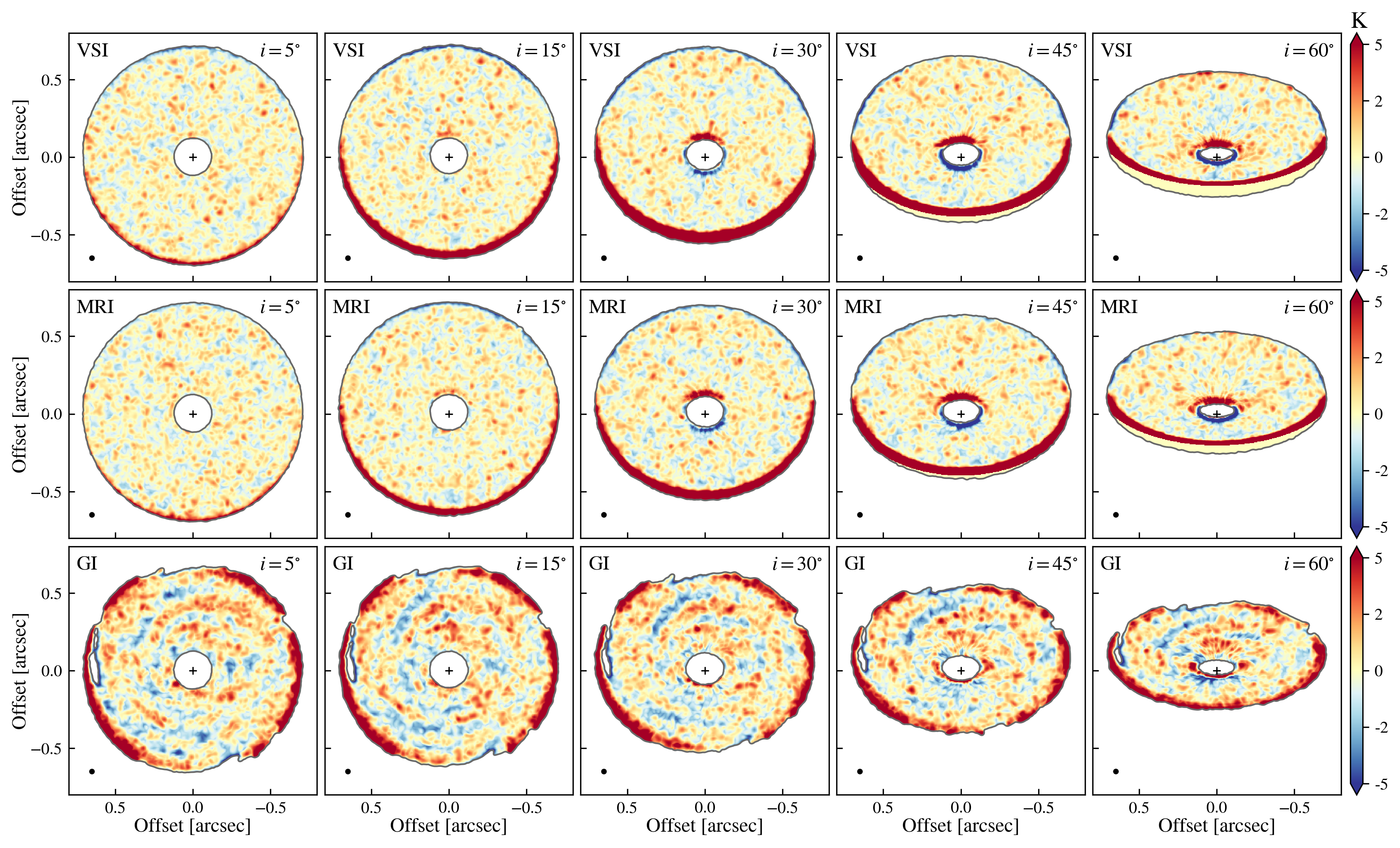}
\caption{Temperature residuals from $\rm ^{12}CO(3-2)$ predictions of inclined disks unstable to VSI, MRI and GI. The residuals are extracted by computing the disks' eighth Moment maps and then subtracting their azimutal averages using \textsc{GoFish}. The models only consider the upper (or front) CO layer. The images have a 104 m s$^{-1}$ velocity resolution, a 35 mas angular resolution, and include the noise level achieved by exoALMA. From left to right, different disk inclinations are shown. The beam size is shown at the bottom left corner of each panel.}
\label{fig:tempresiduals}
\end{figure}

\begin{figure}[htp]
\centering
\includegraphics[angle=0,width=\linewidth]{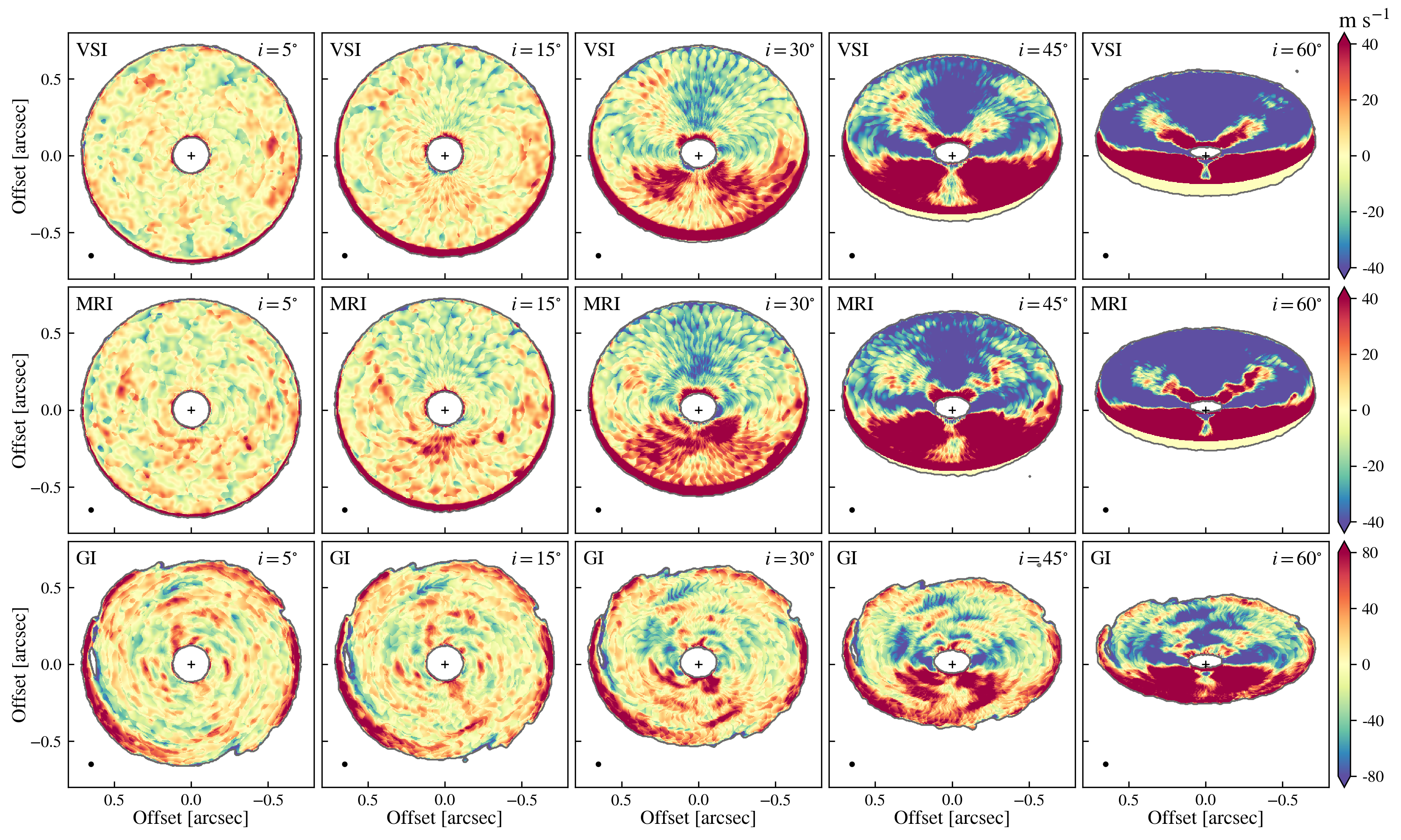}
\caption{Line width residuals from $\rm ^{12}CO(3-2)$ predictions of inclined disks unstable to VSI, MRI and GI. The residuals are extracted by computing the disks' line width maps using the percentiles method in \textsc{eddy}, and then subtracting their azimutal averages using \textsc{GoFish}. The models only consider the upper (or front) CO layer. The images have a 104 m s$^{-1}$ velocity resolution, a 35 mas angular resolution, and include the noise level achieved by exoALMA. From left to right, different disk inclinations are shown. The beam size is shown at the bottom left corner of each panel. The color bar limits have been adjusted for each instability.}
\label{fig:widthresiduals}
\end{figure}

\end{document}